\documentclass[conference]{IEEEtran}

\usepackage{tikz}
\usepackage{amsmath,amssymb,amsfonts}

\usepackage{filecontents}
\usepackage{listings}
\usepackage{caption}
\usepackage{graphicx}
\usepackage{color}
\usepackage{makecell}
\usepackage{multirow}
\usepackage{bbding}
\usepackage{url}
\usepackage{float}
\usepackage{booktabs}
\usepackage{amsfonts}
\usepackage[numbers]{natbib}
\usepackage{longtable}
\usepackage{threeparttable}
\usepackage{subfigure}
\usepackage{hyperref}
\usepackage[ruled,linesnumbered]{algorithm2e}
\usepackage{bm}
\usepackage{wrapfig}
\usepackage[misc]{ifsym}

\newcommand{\projname}{VIM}
\newcommand{\shuffle}{VisualMixer}
\newcommand{\optimizer}{ST-Adam}
\newcommand{\WS}{\text{\textit{WS}}}
\newcommand{\VFE}{\text{\textit{VFE}}}

\newcommand{\minor}[1]{{{\color{black}{#1}}}}
\newcommand\added{\textcolor{black}}{}

\newcommand{\mitt}[1]{ \text{\textit{#1}} }

\newcommand{\ie}{{\em i.e.}}
\newcommand{\eg}{{\em e.g.}}

\begin{document}

\title{You Can Use But Cannot Recognize: Preserving Visual Privacy in Deep Neural Networks\vspace{-0.8cm}}

\author{Qiushi Li\textsuperscript{\dag}*, Yan Zhang\textsuperscript{\dag}*, Ju Ren\textsuperscript{\Letter}*\textsuperscript{\ddag}, Qi Li\textsuperscript{\S}\textsuperscript{\ddag}, Yaoxue Zhang*\textsuperscript{\ddag}\\ * Department of Computer Science and Technology, Tsinghua University, Beijing, China\\\textsuperscript{\ddag} Zhongguancun Laboratory, Beijing, China\\\textsuperscript{\S} Institute for Network Sciences and Cyberspace, Tsinghua University, Beijing, China\\
\{lqs@, yan-zhan23@mails., renju@, qli01@, zhangyx@\}tsinghua.edu.cn}



\IEEEoverridecommandlockouts
\makeatletter\def\@IEEEpubidpullup{6.5\baselineskip}\makeatother
\IEEEpubid{\parbox{\columnwidth}{
    Network and Distributed System Security (NDSS) Symposium 2024\\
    26 February - 1 March 2024, San Diego, CA, USA\\
    ISBN 1-891562-93-2\\
    https://dx.doi.org/10.14722/ndss.2024.241361\\
    www.ndss-symposium.org
}
\hspace{\columnsep}\makebox[\columnwidth]{}}

\maketitle

\def\thefootnote{\textsuperscript{\dag}}\footnotetext{Both authors contributed equally to this work.}
\def\thefootnote{\textsuperscript{\Letter}}\footnotetext{Corresponding author.}


\begin{abstract}
Image data have been extensively used in Deep Neural Network (DNN) tasks in various scenarios, e.g., autonomous driving and medical image analysis, which incurs significant privacy concerns. Existing privacy protection techniques are unable to efficiently protect such data. For example, Differential Privacy (DP) that is an emerging technique protects data with strong privacy guarantee cannot effectively protect visual features of exposed image dataset. In this paper, we propose a novel privacy-preserving framework \textit{\shuffle{}} that protects the training data of visual DNN tasks by pixel shuffling, while not injecting any noises. \textit{\shuffle{}} utilizes a new privacy metric called Visual Feature Entropy (VFE) to effectively quantify the visual features of an image from both biological and machine vision aspects. In \textit{\shuffle{}}, we devise a task-agnostic image obfuscation method to protect the visual privacy of data for DNN training and inference. For each image, it determines regions for pixel shuffling in the image and the sizes of these regions according to the desired VFE. It shuffles pixels both in the spatial domain and in the chromatic channel space in the regions without injecting noises so that it can prevent visual features from being discerned and recognized, while incurring negligible accuracy loss. Extensive experiments on real-world datasets demonstrate that \textit{\shuffle{}} can effectively preserve the visual privacy with negligible accuracy loss, i.e., at average 2.35 percentage points of model accuracy loss, and almost no performance degradation on model training.
\end{abstract}

\section{Introduction}

Neural network models have been 
applied to a wide range of promising image applications, e.g., computer vision, autonomous driving, and medical image analysis~\cite{voulodimos2018deep}. Existing studies~\cite{haim2022reconstructing,geiping2020inverting,zhu2019deep,zhao2020idlg} show that these neural network models can leak the training datasets, \eg, by constructing model reconstruction attacks, \ie, reconstructing training data according to the model weights, gradients, and other model information. However, image data that are used to train these models often contains personal privacy information, such as facial characteristics, license plate numbers, and geographic locations. 
Similarly, medical image data used for training models also involves a large amount of sensitive patient information. 

\begin{figure}[t]
    \center{\includegraphics[width=1\linewidth] {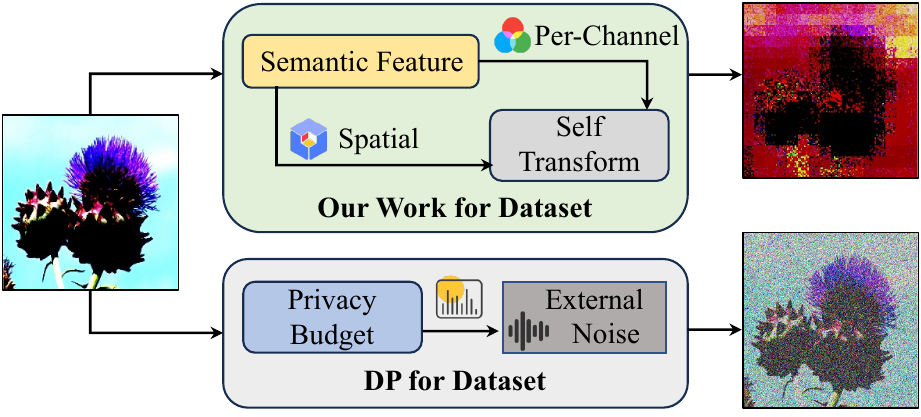}}
    \caption{Our work attempts to protect visual privacy through self-transformation guided by metric of semantic features. DP adds external noise to images, preventing adversaries from distinguishing whether a sample is present in the dataset. This approach is not intended for visual privacy protection in the context of dataset publication.}
    \label{fig: intro}
\end{figure}

A number of privacy-preserving techniques~\cite{boulemtafes2020review} have been developed to address the privacy issues above. For example, homomorphic encryption (HE) enables rigorous data privacy guarantees by encrypting data and ensures that data remains usable yet invisible. 
Trusted Execution Environments (TEEs) based methods utilize trusted hardware to protect data for training DNNs. However, these methods requires substantial extra computational resources or requires specific hardware~\cite{chen2020training} and thus their practicality for image tasks is not clear in practice.
Differential Privacy (DP) is a promising technique that can effectively protect data membership with privacy guarantee. 
By adding controlled noises from predetermined distributions of datasets, DP incurs negligible training delays with acceptable inference accuracy.  
However, as shown in \autoref{fig: intro}, DP is unable to effectively protect visual features of exposed image data because the generated noises that are in high frequency domain can be filtered by human eyes~\cite{wen2022identitydp}. Moreover, DP may incur the obvious performance decrease if the privacy guarantee is strong enough to protect visual feature~\cite{wu2016compiler}. 
Thus, 
it is crucial to protect the privacy of these image data during model training, while maintaining the model performance. 

In this paper, we propose a privacy-preserving framework \textit{\shuffle{}} that aims to protect the training data of visual DNN tasks by shuffling pixels without injecting any noises, which ensures that \textit{\shuffle{}} achieves data protection while retaining the performance of the DNN tasks. 
%
Our framework utilizes a new metric called \textit{Visual Feature Entropy} (VFE) to effectively quantify the visual features of an image from both biological and machine vision aspects. It can accurately evaluate task-agnostic visual features of \minor{an image} with a collection of pixel chromatic value gradients. We can evaluate the uncertainty of visual features in an image by measuring VFE so that we can reduce the amount of privacy information contained in the image. 
In order to achieve desired VFE, we devise \shuffle{}, a task-agnostic image obfuscation method to protect the visual privacy of data for DNN training and inference.
As shown in \autoref{fig: intro}, for each image in the dataset, \shuffle{} decides regions for pixel shuffling in the image and the sizes of these regions according to the desired VFE, and then shuffles pixels both in the spatial domain and in the chromatic channel space in the regions. By shuffling pixels within the determined regions, it can prevent visual features from being discerned and recognized, while incurring negligible accuracy loss. 

Furthermore, since image obfuscation may incur abrupt gradient changes in the global feature space and significant weight gradient fluctuations, it may impede the normal convergence process of the model. In \textit{\shuffle{}}, to address these issues incurred by image obfuscation, we develop an optimizer algorithm called \textit{Adaptive Momentum Stochastic Gradient Descent} (i.e., \optimizer{}). \optimizer{} dynamically adjusts update momentum based on the current gradient and historical gradients to accelerate the model convergence speed and ensure the stability of model training.

We summarize the contributions of the paper as follows. 
\begin{itemize}
\item We propose the first DNN image data protection framework that can effectively protect image data while retaining the accuracy of DNN tasks. 
\item We develop a new metric, Visual Feature Entropy (VFE), which is an effective indicator to measure the visual features of an image and can  quantify the visual privacy of an image. 
\item We propose \shuffle{}, a noise-free and task-agnostic image obfuscation method, to protect the visual privacy of image data in DNN training and inference according to the desired VFE, and devise an optimizer, \optimizer{}, to accelerate the model training performance over the obfuscated iamges.
\item We provide a comprehensive analysis to demonstrate the effectiveness of \shuffle{} and the correlation between the deviation of model outputs and the sizes of shuffling regions. 
\item 
Extensive experiments on real-world datasets demonstrate that \shuffle{} can effectively preserve the visual privacy with negligible accuracy loss, i.e.,  at average 2.35 percentage points of model accuracy loss, and almost no performance degradation on model training.  

\end{itemize}

\section{Threat Model and Design Goals}

\subsection{Threat Model}
\label{sec: threat model}

We consider a general scenario of image data protection in DNN tasks, where clients send their image data (plain or obfuscated) to the server for training a DNN model and then use the trained model for inference. The server 
honestly perform the tasks but tries to steal the corresponding data, \ie, manually identifying images and utilizing attack methods to extract features or information from the DNN model trained by data uploaded from clients.  
Specifically, we focus on the following three attacks. 


\minor{\noindent \textit{Accessing Data uploaded by Clients}. Adversaries on servers can directly access data uploaded by clients. Even when clients obfuscate their images, adversaries still try to recover the original visual features using brute-force or heuristic attacks.}

\minor{\noindent \textit{Reconstructing DNN Trained Data}. Adversaries can perform membership inference attacks on the trained DNN model to identify the ownership of the data used in the training process~\cite{shokri2017membership}. Additionally, they can leverage data augmentation methods such as GAN-based data reconstruction to generate the visual features and recover partial information of the private training data based on the model weights trained by uploaded data from clients~\cite{hussein2020image}.}

\noindent \textit{Recovering Intermediate Gradients and Features}. Adversaries reconstruct visually distinguishable images based on the intermediate gradients~\cite{invisible_fl} and feature maps during the training and inference process~\cite{qiushi_icdcs}.

It is noticed that the label of the image is not private, because it is necessary for training utility.

\subsection{Design Goals}
\label{sec: challenges}
The goal of the paper is to preserve the visual privacy of client image data during the model training and inference process, while retaining the accuracy and performance of the DNN models. The design goals can be summarized as follows.

\noindent \textbf{Quantifying the Task-Agnostic Visual Privacy}.
In the domain of DNN-based vision tasks, the majority of models lack specific interpretability, leading to many visual features involved in the learning process. How to obfuscate the visual features to strike a trade-off between utility and privacy is still an open challenge. Therefore, we should first design a metric to quantify the task-agnostic privacy level of the obfuscated visual features.



\noindent \textbf{Visual-Semantic Obfuscation}.
Generic data encryption methods, provide strict security guarantees but come with significant computational overhead, especially for high-dimension image data.  Unlike traditional data-level obfuscation methods, such as DP that injects noises with a privacy budget, image obfuscation should consider the semantic information of the visual features, to prevent adversaries from obtaining meaningful visual privacy information while keep the data utility for diverse vision tasks. Thus, the second goal is to design a visual-semantic image obfuscation method to balance the model accuracy and privacy level. 

\noindent \textbf{Optimization for Gradient Oscillation}.
Data obfuscation often causes the gradient oscillation problem, making the DNN model  difficult to converge due to the increased randomness of gradients. Existing gradient descent strategies fail to keep the model stability and convergence speed over the obfuscated data. A tailored optimizer should be designed, associate with the data obfuscation method, to tackle the gradient oscillation problem.

\section{Visual Feature Entropy: A New Metric to Measure Visual Privacy}
\label{sec: problem formulation}

\subsection{Limitation of DP on Releasing Visual Datasets}
\label{sec: Limitation of DP on Releasing Visual Datasets}

Some traditional methods, \eg\ Differential privacy (DP), originally designed for statistical data, are facing significant limitations in the field of visual privacy protection. \citet{dp-image} introduced external noise to the feature map to protect the released model from membership detection, but this approach cannot be applied to protecting the released dataset. \citet{differential-privacy} proposed to protect the released model by adding external noise to the dataset, but it fails to protect the visual privacy of image data. As shown in \autoref{fig: intro}, the derived dataset can still reveal some visual information.

The fundamental reason for the insufficient privacy of DP for vision dataset lies in the lack of metrics that can measure the visual privacy. When adding external noise in DP, only the privacy budget is taken into account, rather than the visual features of the images themselves. This results in the inability to eliminate visual privacy from the dataset, even if DP introduces excessive noises that significantly reduce the data utility. Therefore, to eliminate privacy features from the dataset, it is critical to first define a new metric that can measure the visual privacy.

Besides, the privacy metric should be task-agnostic. This is because a dataset usually exhibits multitasking characteristics, where the features that are not privacy-sensitive in one task context may become privacy-sensitive in another. For instance, in an image dataset for autonomous driving,  license plates and pedestrians might expose sensitive privacy, while non-task-related information such as weather and architectural style can still reveal privacy-related details like geographic location. Therefore, in order to eliminate a sufficient amount of visual privacy, we introduce Visual Feature Entropy (VFE) to measure the visual features, ensuring the ability to assess privacy in multitasking scenarios.


\subsection{Definition of Visual Feature Entropy}

In information theory, entropy measures the expected (\ie, average) amount of information conveyed by identifying the outcome of a random trial. The entropy of a random variable is the average level of ``uncertainty'' inherent to the variable's possible outcomes. Inspired by that, we propose \textbf{Visual Feature Entropy} (VFE) to quantify the ``uncertainty'' of visual features. \minor{For an image, a higher VFE exhibits more disorder, indicating a higher uncertainty in visual features, resulting in fewer visual features that can be accurately discerned by biological vision.} With this intuition, the definition of VFE is as follows.

For a given image $I$, we denote the RGB value of a pixel at an arbitrary position $(x,y)$ as $I(x,y), x\in\{0, 1,\dots,N_1-1\}, y\in\{0,1,\dots,N_2-1\}$, where $N_1$ and $N_2$ represents the width and height of an image, respectively.  Consequently, $I(\cdot)$ can be regarded as a discrete function defined over a discrete domain. The ``gradient" of $I(\cdot)$ is defined as
\begin{equation}
    \begin{aligned}
        \nabla_x I(x,y) &= I(x+1, y)-I(x, y),\quad x\in\{0,1,\dots,N_1-1\}\\
        \nabla_y I(x,y) &= I(x, y+1)-I(x, y),\quad y\in\{0,1,\dots,N_2-1\}
    \end{aligned}\label{eq: gradient to subtraction}
\end{equation}
The definition of VFE can be used to measure the visual privacy of any region in an image. For any region $R_I$ with width $w$ and height $h$, if we denote the location of the left top pixel of the region as $(x_0, y_0)$, then the VFE of this region can be calculated as
\begin{equation}
    \VFE_R(R_I) = \sum_{x=x_0}^{x_0+w-1} \sum_{y=y_0}^{y_0+h-1} \left (\nabla_x I(x,y)^2 + \nabla_y I(x,y)^2\right )\label{eq: vfe of R_I}
\end{equation}
Then, the set $\bm{R_I}$ represents the collection of all sub-regions $R_I$ of the image $I$. The VFE of $I$ can be calculated as
\begin{equation}\label{eq: VFE_w}
    \VFE(I) = \frac{F}{N_1N_2}\sum^{R_I\in\bm{R_I}}\VFE_R(R_I)
\end{equation} 
where $F$ is a constant scaling factor (typically set to 1) used to prevent the VFE from being too small or too large value. \minor{The aforementioned definition pertains to the VFE of a single channel, \ie, $\VFE(I)$. For a multi-channel image, the VFE is defined as \begin{equation}
\VFE_M(I) = \frac{1}{C}\sum^{I_c\in I} \VFE(I_c)
    \label{eq: vfe on rgb}
\end{equation}, where $C$ is number of channels, $I_c$ is one channel of the image $I$, and $\VFE_M(I)$ is the VFE of a multi-channel image. In this context, the VFE of a multi-channel image is the average of the VFEs across all channels.}

Under this definition, an image with a higher visual feature entropy indicates that there is a higher density of gradient variations in the image. It also means that the visual features have more ``uncertainty'', are more difficult to identify and hence the image has a higher visual privacy.

\minor{\subsection{Intuition of Visual Feature Entropy}

Fourier transform can provide information in the frequency domain, and it appears to be a potential tool for quantifying this uncertainty and disorder. However, regions of images with frequent gradient changes may not necessarily yield high-frequency information. For instance, multiple low-frequency signals with phase differences can be expressed as high VFE, but Fourier transform only displays as multiple low-frequency signals.

In fact, characterizing this uncertainty using the gradient of the image might be more appropriate. This is because the gradient represents only the absolute difference between adjacent pixels in an image, without reflecting its frequency content~\cite{guo2008spatio}. This implies that we can compute the sum of image variations from the differences between neighboring pixels using the image gradient, thereby avoiding the misleading effects of low-frequency signals. Additionally, by calculating the mean gradient, we can derive a measure of uncertainty akin to what high-frequency signals convey, resulting in an actual representation of VFE. Consequently, the definition of VFE is primarily designed by insights from the image gradient.}

\subsection{How VFE Reflects the Privacy of Visual Features?}
\label{sec: vfe-shuffle-acc}

\begin{figure}[t]
\center{\includegraphics[width=\linewidth] {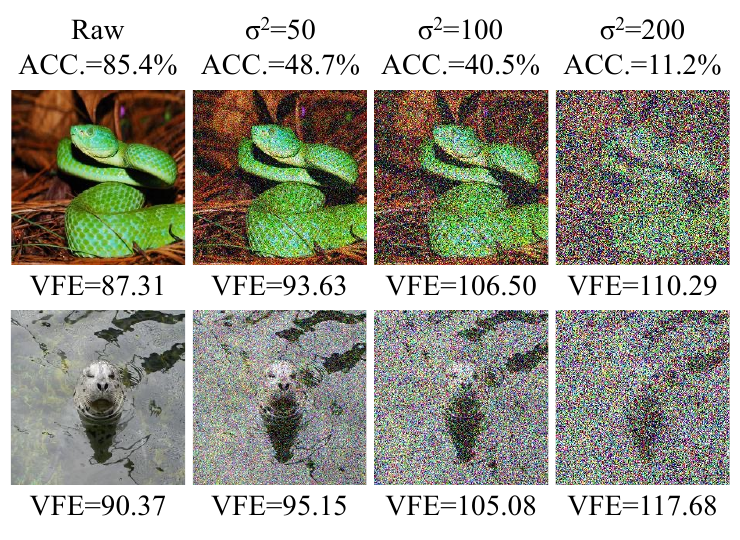}}
    \caption{VFE of Obfuscated Images by Adding More Noises with DP. ($\sigma^2$ reflects the amount of noises added to the image dataset and feature map during training. ACC denotes the accuracy of the ShuffleNet model that is trained using the obfuscated images.)}
    \label{fig: acc-and-vfe-on-dp}
\end{figure}

VFE is an universal metric that can quantify semantically irrelevant visual features for an image. In other words, the methods capable of obfuscating visual features, including Differential Privacy (DP), should enhance VFE. In comparison to the image by DP in \autoref{fig: intro}, we add more Gaussian noise, following DP's rules, to the image dataset to perform actual visual obfuscation. It can be seen from \autoref{fig: acc-and-vfe-on-dp} that, as the adding noise increases, the VFEs of the images are increasing correspondingly and the visual features are becoming more difficult to identify.


However, additional noises on the dataset and feature map may significantly reduce the accuracy of DNN model. Since DP introduces external Gaussian noise that is typically unrelated to the underlying data, it leads to the possibility of some noise occupying a portion of the visual feature space, resulting in a decrease in accuracy. As shown in \autoref{fig: acc-and-vfe-on-dp}, after applying DP and introducing external noise, the accuracy of the ShuffleNet model decreases dramatically. Particularly, when we add noise of $\sigma^2=100$, some detail visual features are not recognizable, such as texture on leaves, but the accuracy of the ShuffleNet model drops to 40.5\% making the dataset useless.

\begin{figure}[t]
    \center{\includegraphics[width=\linewidth] {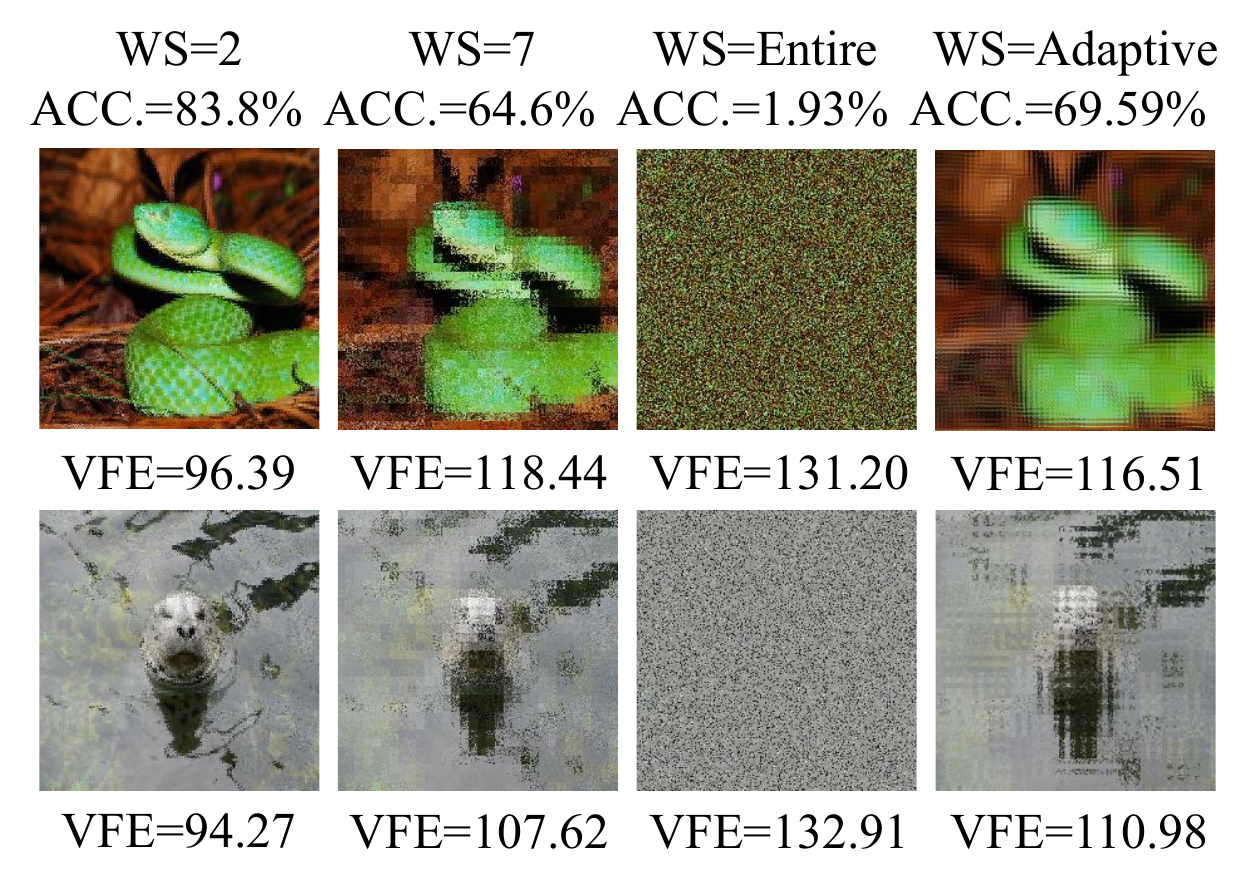}}
    \caption{VFE of Obfuscated Images under Different Shuffling Strategies. ($\WS$ means the window size we used in the corresponding shuffling strategy. ACC denotes the accuracy of the ShuffleNet model that is trained using the obfuscated images.)}
    \label{fig: naive-shuffle-WS}
\end{figure}

Given that adding external noise may potentially lead to a decrease in accuracy, an alternative way to increase the VFE is to spatially shuffle the pixels of the image. However, there are different shuffling strategies to achieve different privacy protection strengths. The extreme way is to randomly shuffle all the pixels of the entire image. It can bring the maximum strength of visual privacy protection, because all the visual features are destroyed entirely, as shown in \autoref{fig: naive-shuffle-WS} (c). In this case, the VFE naturally increases to a very large value but the model accuracy also drops to be zero. 

There is also a soft way to shuffle the images. We can first divide an image to multiple disjoint windows, where each window has equal width and height, denoted by window size $\WS$.  For instance, if $\WS = 7$, it means that we divide the image into a number of windows, each of which contains $7 \times  7$ pixels. Then, in each window, we randomly shuffle the pixels. It is obvious that, as the window size increases, pixel shuffling will break more structural information in the image and lead to an enhanced privacy protection. We can also find from \autoref{fig: naive-shuffle-WS} that the VFE is increasing with the increment of $\WS$, while the model accuracy goes in an opposite direction.

Comparing \autoref{fig: acc-and-vfe-on-dp} and \autoref{fig: naive-shuffle-WS}, we can find that VFE accurately identifies the protection strength of visual privacy under different obfuscation methods. Although both of the methods are facing the dilemma between privacy and data utility, shuffling performs much better than adding-noise in preserving the visual privacy and keeping the data utility. However, to meet the desired privacy requirement, the utility of the obfuscated images, i.e., the accuracy of the DNN model trained by the obfuscated images, is far from satisfaction under such a simple shuffling strategy. In the next subsection, we will discuss how to use VFE to guide the design of a sophisticated shuffling strategy that may achieve the trade-off between privacy and model accuracy. 

\subsection{How VFE Guides the Design of Shuffling Strategy?}

The above section demonstrates that the value of window size, \ie, $\WS$, determines the trade-off between privacy level and data utility when we adopt pixel shuffling to protect the visual privacy. \minor{In each window, besides spatially shuffling the RGB pixels in the image, we can additionally perform per-channel shuffling, \ie, shuffle elements in each channel to increase the VFE and hence to improve the protection of visual features. Per-channel shuffling can effectively obfuscate the color features of an image to preserve the visual privacy but has little impact on the structure and texture features that are more important for the data utility of various vision tasks.}

\minor{In addition to adding a new shuffling dimension, we should focus on why the image obfuscation methods used in \autoref{fig: acc-and-vfe-on-dp} and \autoref{fig: naive-shuffle-WS} experience dramatic accuracy drop.} The main reason is that both of them adopt a uniform data obfuscation strategy in the whole image, without considering the semantic information. It consequently causes that the regions of interests (RoIs) are obfuscated too much to keep the data utility.
According to the definition of VFE, different regions of an image usually have various VFE values. This follows a common sense that different parts of an image contain different amount of visual feature information. With the guidance of VFE, we can design a non-uniform shuffling strategy, where the regions with low VFE should be shuffled under a larger window size while the regions with high VFE can be shuffled under a smaller window size. In such a way, we can increase the VFE of the image to enhance the protection strength on visual privacy and keep more structure and texture information of RoIs.

Based on the design principle, we performed some experiments with a non-uniform shuffling strategy, where we set $\WS=8$ for the windows with original VFE less than average VFE and set $\WS=2$ for the windows with larger VFE. As shown in \autoref{fig: naive-shuffle-WS} (d), we can find that the non-uniform shuffling strategy can improve the model accuracy under a similar VFE.

Thus, it is crucial to design a VFE-guided non-uniform image shuffling strategy to strike the trade-off between visual privacy and data utility. The strategy design should answer the following question: how to divide an image to a number of disjoint windows with non-uniform sizes? Then, the key problem comes to determining the optimal window sizes for different regions in an image to maximize the model accuracy under a given visual privacy protection requirement (i.e., the VFEs of the shuffled images are larger than a specific value).

\section{VisualMixer (VIM): A VFE Guided Privacy-Preserving Image Shuffling Strategy}
\label{sec: the proposed design} 



This section introduces the details of the image obfuscation method, \shuffle{} (VIsualMixer, VIM), and the tailored DNN training optimizer, \optimizer{}. Guided by the VFE of the images, \shuffle{} determines the optimal window sizes for the shuffling strategy to achieve a trade-off between visual privacy and data utility. The tailored optimizer, ST-Adam, is then proposed to address the gradient oscillation problem caused by \shuffle{}.


\subsection{Approach Overview}
\label{sec: overview}

\begin{figure*}[t]
\centering\includegraphics[width=0.9\linewidth] {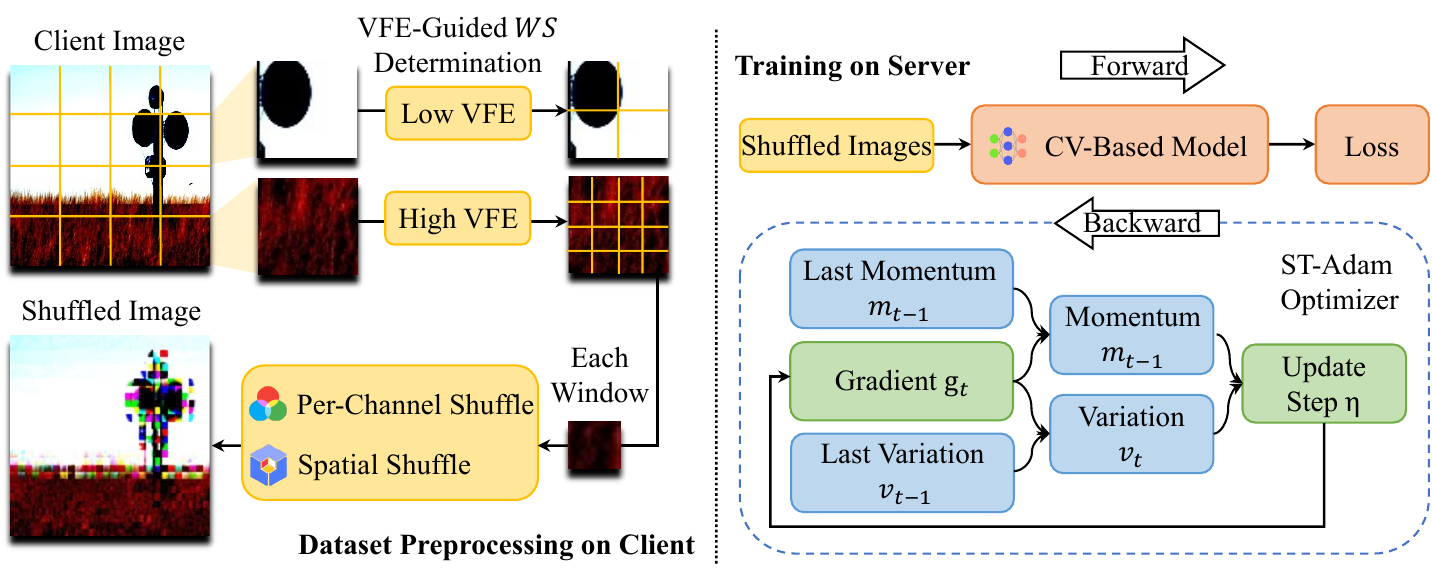}
    \caption{The Architecture and Working Process of \shuffle{}}\label{fig: overall}
\end{figure*}

This section presents the architecture and working flow of \shuffle{}. As shown in \autoref{fig: overall}, \shuffle{} primarily focuses on the data preprocessing stage, aiming to eliminate the visual semantics while preserve the trainable image information. It is guided by the VFEs of different regions in an image to determine the optimal window sizes. In each window, \shuffle{} randomly shuffles the pixels in space and channels to obfuscate the visual features. In addition, since the shuffled images make the training process of the DNN model unstable and very difficult to converge, we design a tailored optimizer, named \optimizer{}, to work with \shuffle{}. By combining momentum optimization and adaptive learning rate adjustment, \optimizer{} can significantly improve the convergence speed of model training over the shuffled image data. 

The working flow of \shuffle{} is as follows. Before the model training stage, clients first use \shuffle{} to obfuscate their images for meeting the privacy requirement. Specifically, \shuffle{} adaptively determines the image shuffling strategy based on the value of VFE and the expectation of model accuracy. Then, clients send shuffled images to server for training. The trained model need work on shuffled images during inference, but do not need decryption.

During the inference stage, we also only need to use \shuffle{} to shuffle the image and then send the shuffled images to the server for inference. It implies that there is no need to modify the model architecture or the underlying computational framework associated with the model itself. Therefore, \shuffle{} can be applied not only in model training and inference scenarios but also in various DNN-related learning tasks.

In the following subsections, we first introduce how to determine the lower bounder of window size to meet the required VFE. Based on that, we illustrate how to determine the optimal window sizes for different regions to maximize the accuracy of a DNN model. Then, we summarize the key step of \shuffle{} into an algorithm, followed by the design details of \optimizer{}.

\subsection{Determining the Lower Bound of Window Size for \shuffle{}}
\label{sec: ws to vfe}

\minor{This section primarily elucidates the relationship between VFE and per-channel shuffling. In particular, it analyzes the mathematical relationship between VFE and the window size on a single channel. This is directly applicable to monochromatic images. For multi-channel images, it remains usability. As inferred from \autoref{eq: vfe on rgb}, the VFE across multiple channels should be the average of the VFEs of each channel, \ie, the effect of shuffling on VFE is additive. Consequently, in the following, we first derive the relationship between VFE and per-channel shuffle.}

In the shuffling strategy, since a larger window size brings a higher VFE, this section first derives the lower bound of window size for \shuffle{}, which can be used to guarantee basic requirement of VFE.

For any region $R_I$ in an image $I$, we denote its window size as $\WS$. Without the loss of generality, we assume that the pixel values in $R_I$ follow a statistically normal distribution. Additionally, after randomly shuffling the pixels within the region, we consider adjacent pixels to be independently and identically distributed.

Based on the assumptions, we can calculate the approximate distribution of the VFE of $R_I$. We denote the pixel values in $R_I$ as $\{p_1, p_2,\dots,p_{\text{\textit{WS}}^2}\}$, then we can obtain the maximum likelihood estimates of their mean and variance:
\begin{equation}
    \mu = \frac{\sum_{x,y}^{R_I} I(x,y)}{\WS^2}, \sigma^2 = \frac{\sum_{x,y}^{R_I}(I(x, y)-\mu)^2}{\WS^2-1}
\end{equation} With the calculated mean $\mu$ and variance $\sigma^2$, we can convert the pixels in $R_I$ into a standard normal distribution 
\begin{equation}
    \hat{I}(x, y) \leftarrow \frac{I(x, y)-\mu}{\sigma}\label{eq: I to I hat}
\end{equation} Now, $\hat{I}(x, y)$ follows a standard normal distribution independently. Therefore, for any two different pixels, \ie, $x_1 \neq x_2 \text{ or } y_1 \neq y_2$, we have 
\begin{equation}
\footnotesize
    \begin{aligned}
    \hat{I}(x_1, y_1)-\hat{I}(x_2, y_2)\sim N(0,2)\Rightarrow \frac{\sqrt{2}}{2}(\hat{I}(x_1, y_1)-\hat{I}(x_2, y_2))\sim N(0,1)\label{eq: p-p normal}
    \end{aligned}
\end{equation}
We need to calculate the distribution of VFE of $R_I$, \ie, $\hat{\VFE}(R_I)$. We substitute \autoref{eq: gradient to subtraction} and \autoref{eq: I to I hat} into \autoref{eq: vfe of R_I}, and get
\begin{equation}
\footnotesize
\begin{aligned}
    \VFE_R(R_I)=\sum_{x,y}^{R_I}(\sigma^2(\hat{I}(x+1, y)-\hat{I}(x, y))^2+\sigma^2(\hat{I}(x, y+1)-\hat{I}(x, y))^2)
\end{aligned}\label{eq: vfe to I hat}
\end{equation} By substituting \autoref{eq: p-p normal} to \autoref{eq: vfe to I hat}, applying the computational rules for normal distributions, we can obtain the $\VFE_R(R_I)$'s distribution $\hat{\VFE}(R_I)$
\begin{equation}
\begin{aligned}
    \hat{\VFE}_R(R_I)&\sim\sum_{x,y}^{R_I}(\sigma^2(N(0, 2))^2+\sigma^2(N(0, 2))^2)\\
    &\sim\sigma^2\sum_{x,y}^{R_I}((N(0, 2))^2+(N(0, 2))^2)
\end{aligned}    
\end{equation}
For $R_I$, there are $\WS(\WS-1)$ pairs of $\nabla_x$ and $\WS(\WS-1)$ pairs of $\nabla_y$. By the definition of the chi-square distribution, we can derive
\begin{equation}
\begin{aligned}
    \hat{\VFE}_R(R_I)&\sim2\sigma^2\sum_{x,y}^{R_I}((N(0,1))^2+(N(0,1))^2)\\
    &\sim2\sigma^2\chi^2_{2\WS(\WS-1)}\\
    \frac{1}{2\sigma^2}\hat{\VFE}_R(R_I)&\sim\chi^2_{2\WS(\WS-1)}   
\end{aligned}     
\end{equation}

It implies that $\frac{1}{2\sigma^2}\VFE_R(R_I)$ follows a chi-square distribution $\chi^2_{2\WS(\WS-1)}$. 
Meanwhile, $\VFE_R(R_I)$ of different $R_I$ can be regarded as following independent identical distribution. Therefore, the sum of the $\VFE_R(R_I)$ of an image $I$ follows a $\chi^2$ distribution, as well, sepcifically,
\begin{equation}
\begin{aligned}
\sum\VFE_R(R_I)&\sim\chi^2_{\frac{2wh}{\WS^2}\WS(\WS-1)}\sim\chi^2_{\frac{2wh(\WS-1)}{\WS}}
\end{aligned}\label{eq: sum_vfe}
\end{equation}
where $w$ and $h$ mean the width and height of the image.
The degree of freedom, $\frac{2wh(\WS-1)}{\WS}$ is so large that we can regard the $\chi^2$ distribution as a normal distribution with the mean of $\frac{2wh(\WS-1)}{\WS}$ and the variance of $\frac{4wh(\WS-1)}{\WS}$. According to \autoref{eq: VFE_w}, the distribution of the $\VFE$ of an image can be got from following equations:
\begin{equation}
\begin{aligned}
    \label{eq: vfeimage}
    \frac{wh}{\WS^2}\VFE(I) &\sim N\left(\frac{2wh(\WS-1)}{\WS},\frac{4wh(\WS-1)}{\WS}\right)\\
    \VFE(I) &\sim N\left(2\WS(\WS-1), \frac{4\WS^3(\WS - 1)}{wh}\right)
\end{aligned}
\end{equation}
When the width and height of the image is set to $224$ (for ImageNet-100) and the window size is set to $8$, the mean of $\VFE(I)$ will be $112$; and the variance of $\VFE(I)$ will be $\frac{2}{7}$.

At this point, we have established the probabilistic relationship between $\WS$ and VFE. Based on this probability relationship, for each image obfuscation, we can compute lower bound of $\WS$ of the ideal VFE with offline manner. This lower bound of obfuscation strength ensures the privacy preservation during the actual obfuscation process.


\subsection{Determining the Upper Bound of Window Size for \shuffle{}}
\label{sec: diff_max to ws}

In \autoref{sec: ws to vfe}, we present that the lower bound of window size $\WS_l$ is constrained by the desired VFE. Since the increment of window size leads to a decrease in data utility, this section proceeds to derive an upper bound of $\WS$ that can optimize the accuracy of the DNN model. Based on the two bounds, we can attain the optimal $\WS$ for \shuffle{} to calibrate the privacy-utility trade-off correctly.

The challenge of optimizing the data utility of \shuffle{} lies in quantifying the impact of shuffling on the accuracy of a DNN model under given window sizes. Instead of directly measuring the model accuracy, we use the output bias of the model to represent the impact of shuffling on data utility. The idea is that, for a specific shuffling strategy with a given $\WS$, we calculate the maximum output bias between the original image and the shuffled image. Then, we use the maximum output bias as the indicator of data utility to determine the upper bound of window size $\WS_u$. In what follows, we use mathematical induction to perform our analysis.


Currently, in the domain of DNN for CV-tasks, the primary architectural choices of are Convolution-based or Transformer-based. Both Transformer and Convolution are linear structures and share fundamental similarities. Therefore, in this section, we provide the calculation proof for $\WS_u$ using the CNN as an example.

For base case in mathematical induction, we begin by considering the typical module of an CNN model, consisting of a convolutional layer followed by a max-pooling layer.
Denote $W$ as the weight of convolutional kernel with a size of $N_{CK}$ (CK represents ConvKernel) and a stride of $N_{CS}$ (CS represents ConvStride), where its parameters $w_{i}$ follows a normal distribution, i.e., $w_{i}\sim N(\mu_W, \sigma_W)$. The elements within $W$ are arranged from largest to smallest as $\{w_1,w_2,\dots,w_{size}\}$. Denote the size of the max-pooling operator as $N_{\mitt{MP}}$ and the size of the stride as $N_{\mitt{MPS}}$, where $N_{\mitt{MP}}=N_{\mitt{MPS}}=2$ is a common setting for max-pooling operators. We set the parameters of the convolutional layer as a minimum $N_{\mitt{CK}}=2\times 2$ and $N_{\mitt{CS}}=1$. 

Consistent with the previous statement, we denote the window size of a region in image $I$ as $\WS$, which is set to $3$ for base case in mathematical induction. We first calculate the output bias in this base case and then generalize it to all cases where $N_{\mitt{CK}}\ge 2\times 2$ and $\WS\ge 3$. We normalize the pixel value in the window, denoted by $I_{ij}\in [0,1]$. $I'$ denotes the shuffled image of $I$; $A$ and $A'$ denote the feature map obtained through the convolution layer in which $I$ and $I'$ has been processed, respectively; $B$ and $B'$ denote the outputs after $A$ and $A'$ are processed by the max-pooling operator, respectively. Our purpose is to calculate the distribution of $\mitt{diff}_{max}$ that denotes the maximum difference between $B$ and $B'$. Since the sizes of $B$ and $B'$ are $1\times 1$, we can get possible  $B_{max}$ and $B_{min}$
\begin{equation}
\footnotesize
    \begin{aligned}
        &B_{max} = \max\{A_{ij}\} \leq \sum I_{ij} \cdot W_{ij}, \text{if }W_{ij} > 0 \text{ then }I_{ij} = 1\text{ else }I_{ij} = 0
    \end{aligned}\label{eq: b_max}
\end{equation}\begin{equation}
\footnotesize
\begin{aligned}
        & B_{min} = \max\{A_{ij}\} \geq \sum I_{ij} \cdot W_{ij}, \text{if }W_{ij} < 0 \text{ then }I_{ij} = 1\text{ else }I_{ij} = 0
    \end{aligned}\label{eq: b_min}
\end{equation} 
Therefore, the maximum output bias can be calculated as 
\begin{equation}
\small
\mitt{diff}_{max}\ =| B -B'| = B_{max}-B_{min}
\end{equation}
By probability theory, the expectation of $\mitt{diff}_{\max}$, i.e., $\mathbb{E}(\mitt{diff}_{\max}) $, can be calculated as an integral over a four-dimensional space $D$ (because of $N_{\mitt{CK}}=2\times 2$)
\begin{equation}
\footnotesize
\begin{aligned}
    &\iiiint_D (w_1+w_2)p(w_1)\frac{p(w_2)}{\Phi(w_1)}\frac{p(w_3)}{\Phi(w_2)}\frac{p(w_4)}{\Phi(w_3)}dw_1dw_2dw_3dw_4\\
    &D = \{\vec{w}=(w_1, w_2, w_3, w_4)|\vec{w}\in\mathbb{R}^4 , w_1\ge w_2\ge w_3\ge w_4\}
\end{aligned}\label{eq:expt}
\end{equation}
where functions $p(\cdot)$ and $\Phi(\cdot)$ represent the density function and the cumulative distribution function, respectively.

However, it is infeasible to derive the closed-form expectation of $\mitt{diff}_{\max}$ by Equation \eqref{eq:expt}. Then, we can give a complete induction to calculate $B$, by enumerating all possible cases of $I$ in extreme value space, \ie, $\{0, 1\}$. With a $3 \times 3$ matrix of $I$, it has totaling $2^{3\times 3}=512$ cases. Based on our assumptions, we know that there could be only five possibilities for the sign combination of $w_1, w_2, w_3, w_4$. Combined with the 512 cases of I, there are a total of 2560 cases. We have enumerated all of these.
\begin{table}[h]
\centering\caption{Complete induction of all possible output $B \text{ in } I \in \{0,1\}$.}
\label{tab: complete induction}
\begin{tabular}{lll}
    \toprule
    \textbf{$B$} & \textbf{Number} & \textbf{Percentage}\\
    \midrule
    0 & 200 & 7.8\%\\
    w1 & 457 & 17.9\%\\
    w2 & 212 & 8.3\%\\
    w3 & 70 & 2.7\%\\
    w4 & 9 & 0.4\%\\
    w1 + w2 & 473 & 18.5\%\\
    w1 + w3 & 247 & 9.6\%\\
    w1 + w4 & 60 & 2.3\%\\
    w2 + w3 & 140 & 5.5\%\\
    w2 + w4 & 15 & 0.6\%\\
    w3 + w4 & 3 & 0.1\%\\
    w1 + w2 + w3 & 411 & 16.1\%\\
    w1 + w2 + w4 & 92 & 3.6\%\\
    w1 + w3 + w4 & 28 & 1.1\%\\
    w2 + w3 + w4 & 5 & 0.2\%\\
    w1 + w2 + w3 + w4 & 138 & 5.4\%\\
    \bottomrule
\end{tabular}
\end{table}


The enumeration data in \autoref{tab: complete induction} allows us to perform Monte Carlo simulations, taking into account the probability of each combination and the assumption that the convolution kernel follows a normal distribution. This allows us to determine the parameter d at any level of confidence. As the size of the convolution kernel increases, similar methods can be used for calculation.




Moreover, to show the influence of the conclusion above given by size increasing of $I$ and $I'$, we define the following notations. $S_I(')$ is the size of $I(')$, similarly for $S_A(')$ and $S_B(')$. Since the stride of maxpooling operator is $2$, we only consider the situation where $\Delta S_I = 2m, m=1,2,\dots.$ When $\Delta S_I = 2m-1$, the maxpooling operator will trigger padding operation, which makes the situation similar to the case where $\Delta S_I = 2m$. Then $\Delta S_A = 2m$ and $\Delta S_B = m$, which means that there is $m^2$ elements in $B$ and $B'$. Since each element of $B$ and $B'$ has independent predecessors in $I$, and $\mitt{diff}_{\max}$ only takes the maximum value, we can apply the multiplication rule of independent event probability calculations here. Therefore, we can obtain $\mathbb{P}(\mitt{diff}_{\max} \leq d)$
. Then the probability that all of the $\mitt{diff}_{\max}$ with respect to all elements in $B$ and $B'$ are less than $d$ should be 
\begin{equation}
    \mathbb{P}(\mitt{diff}_{\max} \leq d) = \alpha^{m^2},~ \mitt{if } N_{CK}\ge 2\times 2 \text{ and } \WS\ge 3
\end{equation}

When the size of $W$ increases, \ie\ $m$ increases, which means the receptive field increases, according to \autoref{eq: b_max} and \autoref{eq: b_min}, $\mitt{diff}_{\max}$ will definitely increase.  It is unnecessary to consider the case where the $S_I$ is larger than $W$, because $\mitt{diff}_{\max}$ is the maximum value obtained after exhausting all possible distributions of $I$. When the $S_I$ is larger than the convolution kernel size, the convolution kernel will perform a sliding window on $S_I$. The size of each window remains the same size as the convolution kernel size during the sliding. Meanwhile, the $\mitt{diff}_{\max}$ obtained after all sliding windows must be smaller than or equal to the $\mitt{diff}_{\max}$ obtained after exhausting all possible distributions in the sliding window. It is worth noting that the mainstream structure of CNN consists of a convolution layer and a batch normalization layer. The job of pooling layers, down-sampling, is done by convolution layers. However, the principle of mixing operation does not change; Thus, the conclusion also works in current CNNs.

Then, we adopt a reverse-solving method to determine the maximum allowable \textit{WS}. The process to calculate the optimal value of \textit{WS} for balancing privacy protection and accuracy loss begins with setting an initial value of \textit{WS}, denoted as \textit{WS$_0$}, to 3. Using a Monte Carlo simulation, we determine the value of $\alpha_0$ that satisfies the condition $P(\mitt{diff}_{max} \leq d) = \alpha_0$. Next, we satisfy $P(\mitt{diff}_{max} \leq d) = \alpha$ by setting a given $\alpha$. By setting $\alpha = \alpha_0^{m^2}$, we can calculate the value of $m$ as $\sqrt{\log_{\alpha_0} \alpha}$. And finally, we compute \textit{WS} by $\textit{WS} = \textit{WS$_0$} + 2 \times \lfloor m \rfloor$. This process dynamically calculates the value of $n$ to achieve a balance between data privacy and inference accuracy.

The shuffling is performed by randomly rearranging the pixels of the image, thereby changing the original feature distribution of the image. Shuffling can be done at the pixel level or at the block level, according to \textit{WS}. Therefore, we first divide the image into sub-regions, and then we calculate the VFE of each sub-region. Then, we compare the VFE value of each sub-region with the median $\VFE_m$. If $\VFE_i > \VFE_m$, we use a larger \textit{WS} for shuffling; otherwise, we use a smaller \textit{WS} for shuffling.

\subsection{Algorithm of \shuffle{}}
\label{sec: algorithm flow}

\begin{algorithm}[t]
    \caption{Image Processing in \shuffle}\label{alg}
    \KwData{Input image $I$}
    \KwOut{Shuffled image $I'$ by \shuffle}
    $\WS_{u}\leftarrow\WS_0 + 2 \times \lfloor \sqrt{log_{\alpha_0}\alpha} \rfloor$ // From \autoref{sec: diff_max to ws} \\
    $\WS_{l}\leftarrow\WS$ from the distributions of target VFE\\
    $\WS\leftarrow Size(I)$\\
    \While{$\WS > \WS_{u}$}{
        $\WS \leftarrow 2^{\left\lfloor \log_2 \WS/2\right\rfloor}    $
    }
    $\bm{R}\leftarrow$ Dividing image $I$ into regions based on $\WS$ \\
    \While{$\bm{R}\neq\varnothing$} {
        $\bm{R}_i\overset{{\scriptscriptstyle\operatorname{R}}}{\leftarrow}\bm{R}$\\
        $\WS\leftarrow Size(\bm{R}_i)$\\
        \uIf{$\WS\leq2^{\left\lfloor \log_2 \WS_{l}\right\rfloor}$}{
                $\WS = 2^{\left\lfloor \log_2 \WS_{l}\right\rfloor}$\\
                Spatial and per-channel shuffle $\bm{R}_i$ with $\WS$\\
                $\bm{R}=\bm{R}-\{\bm{R}_i\}$
        }\Else{
            $\WS = \left\lceil Size(\bm{R}_i)/2\right\rceil$\\
            \uIf{$\VFE(\bm{R}_i)\leq\VFE_{m}$}{
                Spatial and per-channel shuffle $\bm{R}_i$ with $\WS$\\
                $\bm{R}=\bm{R}-\{\bm{R}_i\}$
            }\Else{
                $\bm{R}\leftarrow\bm{R}~\cup$ Dividing $\bm{R}_i$ into regions based on $\WS$
            }
        }
    }
    \Return $I'$
\end{algorithm}

The detailed algorithm of \shuffle{} is summarized in \autoref{alg}. Firstly, we need to calculate the upper and lower bounds of $\WS$ offline, $\WS_{u}$ and $\WS_{l}$. The $\WS_{u}$ is used to control the maximum deviation of model output, \ie\ $\text{diff}_{max}$, thereby preventing excessive loss of model accuracy. Based on the probability relationship proven in \autoref{sec: diff_max to ws}, we can obtain the $\WS_{u}$ for confidence probability $\alpha$ through the reversing method. The $\WS_{l}$ is used to control the expected VFE in order to ensure privacy. According to the conclusion proven in \autoref{sec: ws to vfe}, we know that $\frac{1}{2\sigma^2}\VFE$ follows a chi-square distribution. Therefore, by performing a backward table lookup, we can obtain the $\WS_{l}$ for the target VFE interval.

After obtaining the upper and lower bounds of $\WS$, we need to adapt suitable $\WS$ for different regions of the image based on their VFE. Firstly, we scale down $\WS$ to the nearest power of two that is less than or equal to $\WS_{u}$, ensuring the subsequent iterative $\WS$. Next, we partition the image according to the current $\WS$ and add the segmented images to the set $\bm{R}$ of images to be processed. At each iteration, an image $\bm{R}_i$ is randomly selected from the set $\bm{R}$. Firstly, if its $\WS$ is lower than $\WS_l$, it is directly shuffled using the $\WS$ closest to $\WS_l$, which is a power of 2. If its $\WS$ is between the upper and lower bounds, it is determined based on the VFE. If the VFE is lower than the median VFE of the initial segmented set, it is shuffled using the next level of $\WS$, \ie\ $Size(\bm{R}_i)/2$. If the VFE is greater than the median VFE, it is partitioned into smaller blocks and added to the $\bm{R}$, awaiting the next round of $\WS$ and VFE checks.

This process is convergent, ensuring that all sub-images on the entire image will be protected by shuffling with dynamic $\WS$ intensities between $\WS_l$ and $\WS_u$. It also ensures that the shuffle intensity stays within the required range, not exceeding $\WS_u$ for accuracy and not going below $\WS_l$ for VFE.

\subsection{Stable Adaptive Moment Estimation}
\label{sec: ST-Adam}

\begin{figure}[t]
    \center{\includegraphics[width=1\linewidth] {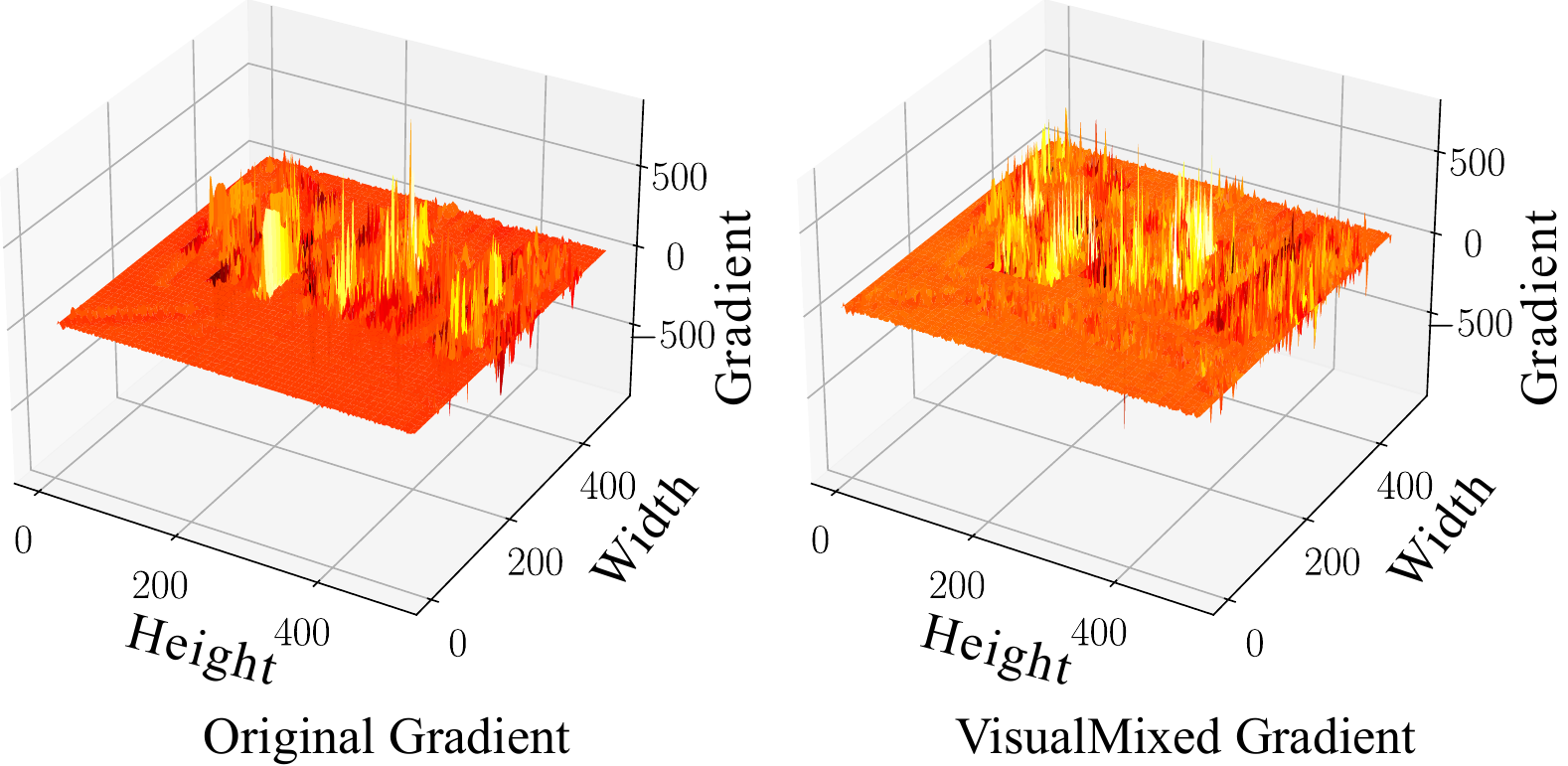}}
    \caption{\minor{Comparing gradients of original and VisualMixed images.}}
    \label{fig: gradients}
\end{figure}

When using Adam to train models on the mixed dataset, we find that it is hard to get an ideal minimum so that the performance of trained models decreases sufficiently. To address the issue of models struggling to converge due to gradient oscillation in certain scenarios involving small datasets and individual models, we propose Stable Adaptive Moment Estimation (ST-Adam). ST-Adam achieves the ability to converge as expected in the presence of gradient oscillation by employing dual adaptivity for both momentum and learning rate. As shown in \autoref{fig: gradients}, when training VGG on CIFAR10, the original gradients exhibit distinct features that facilitate rapid gradient descent for optimizers like Adam, if no data obfuscation is applied. However, when training on obfuscated data, there is a significant fluctuation in gradients, referred to as gradient oscillation. This phenomenon results in a substantial loss in the model's convergence capability.

Here is the description of the update rules of ST-Adam. Firstly, the optimizer calculates the gradient of the loss function based on the value of parameters, $g_t = \nabla f(w_t)$. Then, According to the gradient and pre-defined hyperparameter $\beta$, the optimizer calculate the momentum of the loss function, $m_t = \beta \times m_{t-1} + (1 - \beta) \times g_t$. After that, using another hyperparameter, $\gamma$, it get the adaptive learning rate, $v_t = \gamma \times v_{t-1} + (1 - \gamma) \times g_t^2$. Finally, based on all of the information above, the optimizer can update the parameters of models, $w_{t+1} = w_t - \eta * \frac{m_t}{\sqrt{(v_t)} + \epsilon}$.

Here, $w_t$ represents the weights at timestep $t$, $g_t$ represents the gradient, $m_t$ represents the momentum, $v_t$ represents the adaptive learning rate, $\eta$ is the initial learning rate, $\beta$ and $\gamma$ are the decay coefficients of momentum and adaptive learning rate, respectively, and $\epsilon$ is a smoothing term to prevent the denominator from being $0$. We assume the objective function $f(w)$ is convex. In convex optimization, we are concerned with whether the distance between the optimized objective function value and the optimal solution decreases as the number of iterations increases. We define 
\begin{equation}
    \Delta w_t = w_t - w^* ,\quad \Delta f_t = f(w_t) - f(w^*)
\end{equation}

Here, $w^*$ is the optimal solution. For convex functions, according to Jensen's inequality, we have 
\begin{equation}
    \Delta f_t \leq g_t^T \times \Delta w_t
\end{equation}

Substituting the update rule of ST-Adam, we can obtain 
\begin{equation}
    \Delta f_t \leq (\frac{m_t}{\sqrt{v_t} + \epsilon})^T \times \Delta w_t
\end{equation}

We can observe that when the gradient $g_t$ is large, the momentum $m_t$ and the adaptive learning rate $v_t$ will also increase accordingly. When the gradient is small, $m_t$ and $v_t$ will decrease. This means that in areas with larger gradients, the optimizer will use larger update steps, thus speeding up the convergence process; in areas with smaller gradients, the optimizer will use smaller update steps to maintain a stable optimization process. Therefore, we can conclude that the convergence of ST-Adam is guaranteed. We added weight decay and momentum terms. Weight decay helps prevent model overfitting, especially when encountering drastic and dense gradient distributions during training. The momentum term helps speed up the optimization process, making it easier for the model to converge when encountering drastic gradient changes.

The main difference between the Adam and ST-Adam is that during the calculation of update step, Adam will rescale the step length according to the increase of time; while the step length obtained by ST-Adam is more stable. More specifically, in Adam, we need extra two calculations:
\begin{equation}
    \begin{aligned}
        \hat{m_t} & = m_t/(1-\beta^t)\\
        \hat{v_t} & = v_t/(1-\gamma^t)
    \end{aligned}
\end{equation}
Those two steps rescale the value of $m_t$ and $v_t$, which can make the update step of the Adam more flexible. In most cases, this will be an improvement for an optimizer, for the reason that it make it possible to let optimizer choose the update step adaptively according to the shape of the loss function in the neighbor. However, In the scenario of \shuffle{}, the gradients are more likely to change dramatically within a limited range due to the mixing operation. That is to say, the flexible length of update steps is more like a poison than a benefit for our model training process. Although adaptive update steps con let the model converge more quickly, it makes the model get trapped into a local minimum with higher probability. Therefore, it is better keep the update steps stable to avoid the model being trapped in local minimum. That's why our ST-Adam remove the above two steps.

By introducing momentum and adaptive learning rate adjustments, the ST-Adam optimizer can maintain good convergence performance when dealing with data after \shuffle{}. In \autoref{sec: evaluation}, we applied the ST-Adam optimizer to different convolutional neural network models and compared it with traditional SGD optimizers and other popular optimizers, such as Adam. Experimental results show that the ST-Adam optimizer exhibits faster convergence and higher stability. 

\section{Evaluation}
\label{sec: evaluation}

\subsection{Experimental Setup}
\label{sec: experimental settings} 

\textbf{Testbed and Baselines.} We employ a single NVIDIA Geforce RTX 3090 GPU as testbed. The code is executed on Ubuntu 20.04, using the framework of PyTorch version 1.9. To validate the performance of \shuffle{}, \minor{we evaluate it against three advanced privacy-preserving methods: (a) InstaHide \cite{instahide}, received the 2020 Bell Labs Prize second place award, which is an obfuscation-based approach;} (b) a differential privacy based method; (c) a federated learning method with differential privacy. These comparisons are presented in \autoref{tab: throughput}.

\textbf{Datasets and DNN Models.} \autoref{tab: accuracy loss} present the datasets used in our work. Four representative datasets in the CV domain were selected for testing. \minor{Specifically, ImageNet-100 is a standard public dataset released by the well-known Kaggle contest~\cite{imagenet-100}.} And CIFAR-10 is a color image dataset consisting of 10 categories, with a total of 60,000 images; MNIST is a dataset for handwritten digit recognition, containing 60,000 training samples and 10,000 testing samples; AT\&T dataset, also known as the ORL (Olivetti Research Lab) face database, consists of 400 grayscale face images belonging to 40 different individuals, with 10 images per individual. This dataset is commonly used for face recognition and face detection tasks. The selection criteria for the DNN models used in this work are as follows: (1) the test models should encompass a variety of mainstream frameworks for CV tasks, including Transformer structures~(e.g., ViT-B \cite{vit} and Swin-T \cite{swint}), directly connected CNNs~(e.g., AlexNet \cite{alexnet} and VGG \cite{vgg}), and residual network models (e.g., ResNet \cite{resnet} and DenseNet \cite{densenet}); (2) the DNN models should cover network models with varying parameter scales and computational efforts, including very large networks like ViT-B, large networks like VGG, and lightweight networks like MobileNet~\cite{mobilenet} and ShuffleNet \cite{shufflenet}. 

\subsection{Validation on ST-Adam Optimizer}
In this section, we conducted experiments to validate the performance of the ST-Adam optimizer. We trained models using both ST-Adam and widely used Adam optimizers on the CIFAR10, MNIST, and ImageNet-100 datasets. Throughout the training process, we recorded the accuracy and loss curves of the models for later comparison.

\begin{figure*}[t]
\includegraphics[width=\linewidth]{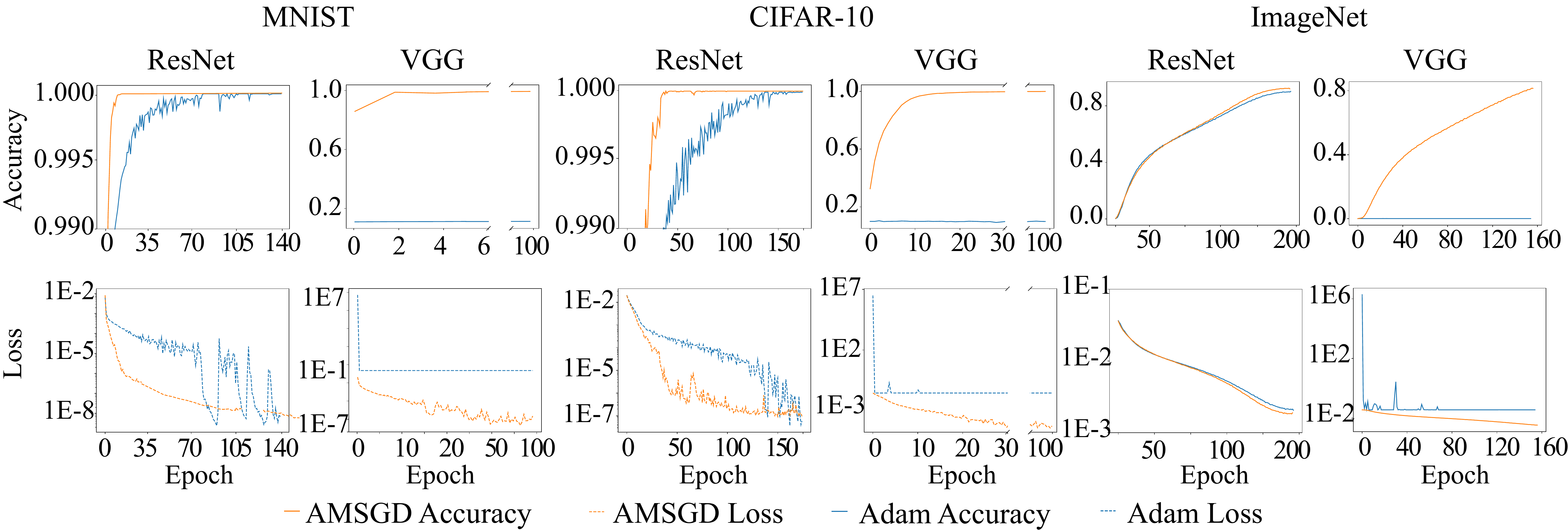}
\caption{The training curves of ST-Adam and Adam optimizers on MNIST, CIFAR-10, ImageNet.}
\label{fig: curves of optimizers}
\end{figure*}

Based on the experimental results in \autoref{fig: curves of optimizers}, the ST-Adam optimizer outperformed the Adam optimizer in terms of both accuracy and loss curves. This indicates that ST-Adam exhibits better optimization performance and faster convergence speed on these datasets. The reasons behind this superiority can be explained from the following three aspects.

\begin{figure}[t]
\vspace{-0.3cm}
    \includegraphics[width=\linewidth]{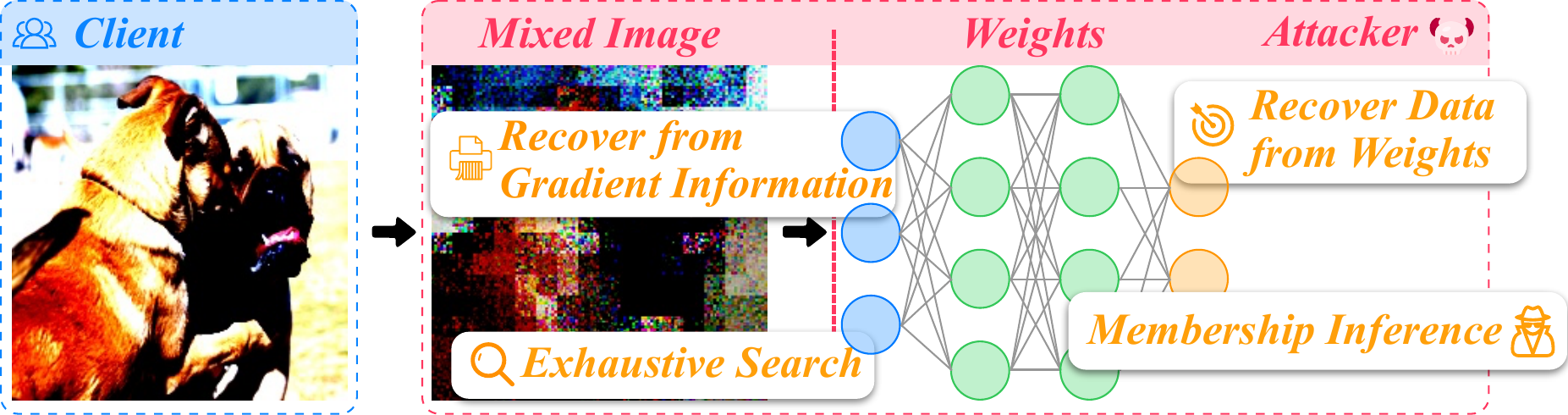}
    \caption{The attack flowchart shows the possible parts that may be attacked: (1) Attacks against the shuffled images, such as exhaustive search restoration and restoration using gradient information; (2) Attacks against the training weights, such as GAN-based attack method \cite{gan-attacks} and membership inference attack method \cite{membership-inference-attacks}.}
    \label{fig: attacks flow}
\end{figure}

\subsection{Defend Against Privacy Leakages}\label{sec: defend against privacy leakages}

\textbf{Attack flow.} \autoref{fig: attacks flow} illustrates the privacy attacks in the current landscape. In typical scenarios, users obfuscate their images before uploading them to the server for model training. In the threat model we defined in \autoref{sec: threat model}, the server is a semi-honest adversary. Therefore, in this process, privacy primarily faces four types of threats.
These attacks include exhaustive search, \ie, brute-force crack, heuristic attacks by shredder recovering algorithm, weights leaking visual feature and membership inference on weights.

\textbf{Exhaustive search.} When attempting to restore \projname{} data through brute force cracking, we can analyze it using the principle of sorting. As explained in \autoref{sec: the proposed design}, the total number of permutations $N$ of exhaustive search is calculated based on the probability ranking principle as
$N = \sum_{i = 0}^{\frac{wh}{\WS^2}^2} (\WS_i)^2!$. With the equation, we observe that when $\exists \WS_i \geq 6$, the total number of permutations $N$ exceeds $2^{128}$, a number that can be reached regularly in our experiments. Thus, it becomes apparent that restoring an \projname{} image solely through brute force methods would necessitate an overwhelmingly large number of attempts, rendering it practically impossible.

\textbf{Heuristic Attacks by Shredder Challenge Algorithm}
Based on the proposed design above, it is evident that restoring images processed by \shuffle{} using brute-force methods is challenging. However, there is a possibility of attempting recovery using gradient information between pixels, which is used in Shredder Challenge Algorithm\cite{shredder}. To test the resilience of \shuffle{}, against this restoration method, we conducted experiments in this section.
\begin{figure}[t]
    \includegraphics[width=0.95\linewidth]{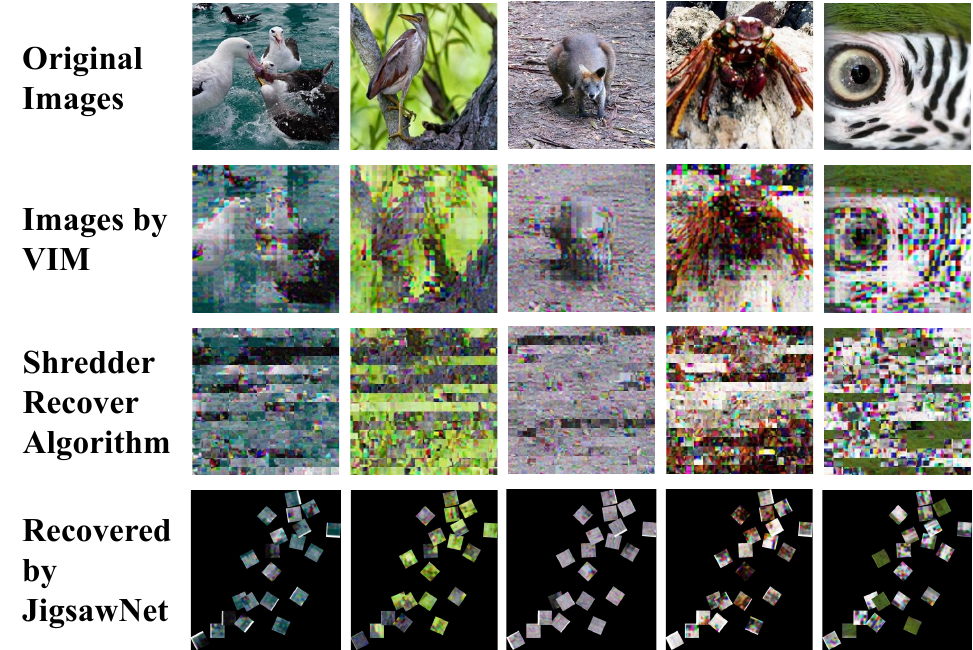}
    \caption{\minor{Restoring results of the obfuscated images using shredder recovering algorithm by gradient information and \cite{jigsawnet}.}}
    \label{fig: gradient attack results}
\end{figure}

The experimental results presented in \autoref{fig: gradient attack results} demonstrate that the  shredder recovering effect is not ideal. \minor{This is primarily due to the \projname{} process employed by \shuffle{}, which alters the relative positions of pixels in both spatial and each channel, thereby disrupting the visual features and structural information of the original image. This means that the relative positions of the pixels in space and its channel have changed, thereby destroying the visual features and information structure of the original image. JigsawNet employs DNNs to construct latent relationships between sub-images, thereby attempting to piece together the original image~\cite{jigsawnet}. It represents another heuristic attack method. However, such approaches typically perform well when the fragment size is larger. In our method, the fragment size is effectively $1$, as we shuffle all pixels within the $\WS$. As illustrated in the \autoref{fig: gradient attack results}, its attack efficacy is limited in our context.}

\textbf{Feature Restoration from Model Weights by GAN.} We attempt to use the method presented in \cite{gan-attacks} to attack the weights from LeNet-like model obtained from training on the raw data (Celeba and MNIST), intermediate updates during federated learning, and training on the dataset processed by differential privacy($\sigma=50$) and our \projname{}. 
\begin{figure}[t]
    \centering\includegraphics[width=0.9\linewidth]{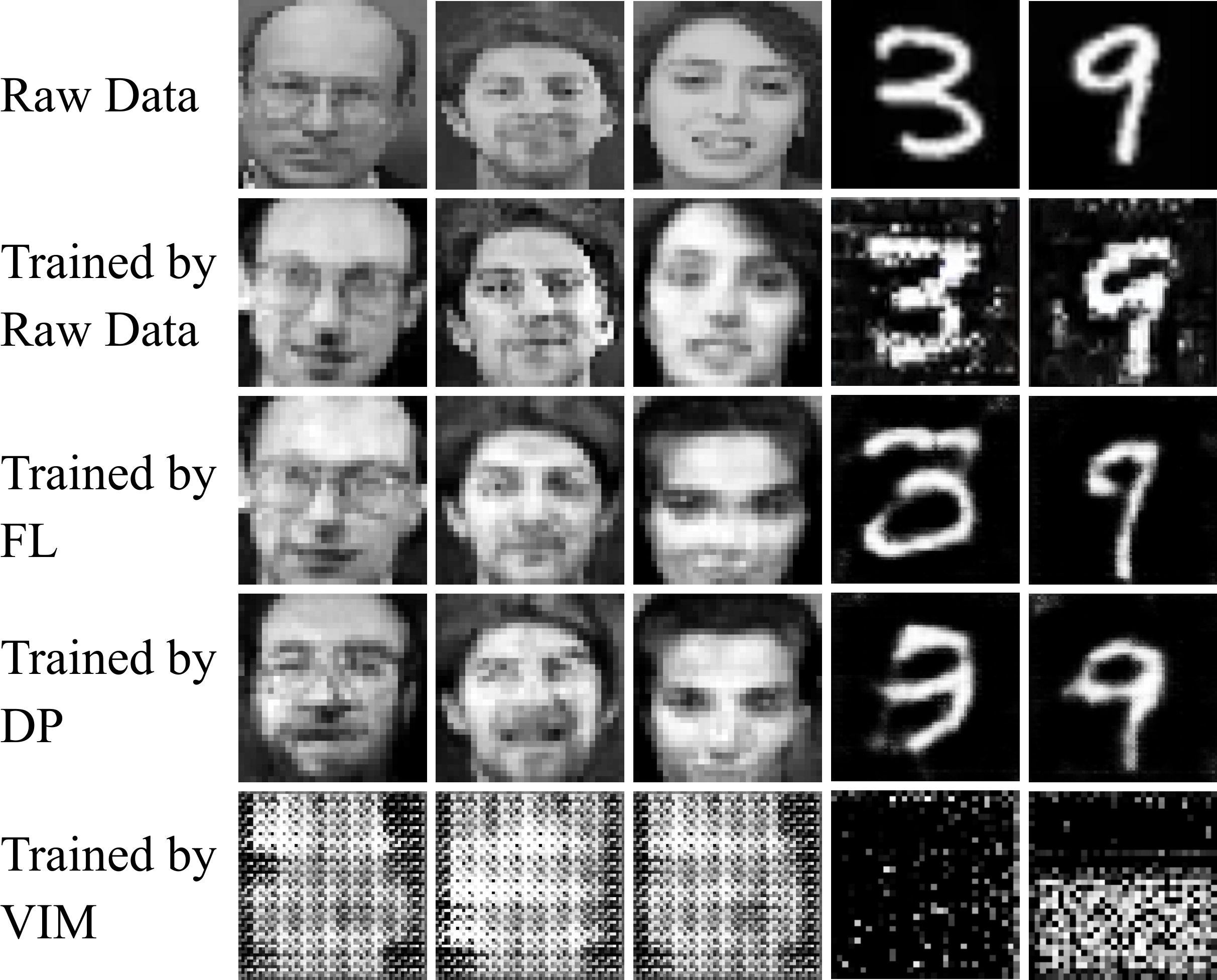}
    \caption{Comparison of the attack effects on the model weights based on the GAN model. The results indicate that these attacks fail to recover any meaningful original information from the protected weights.}
    \label{fig: gan attacks results}
\end{figure}

\begin{figure}[t]
\includegraphics[width=\linewidth]{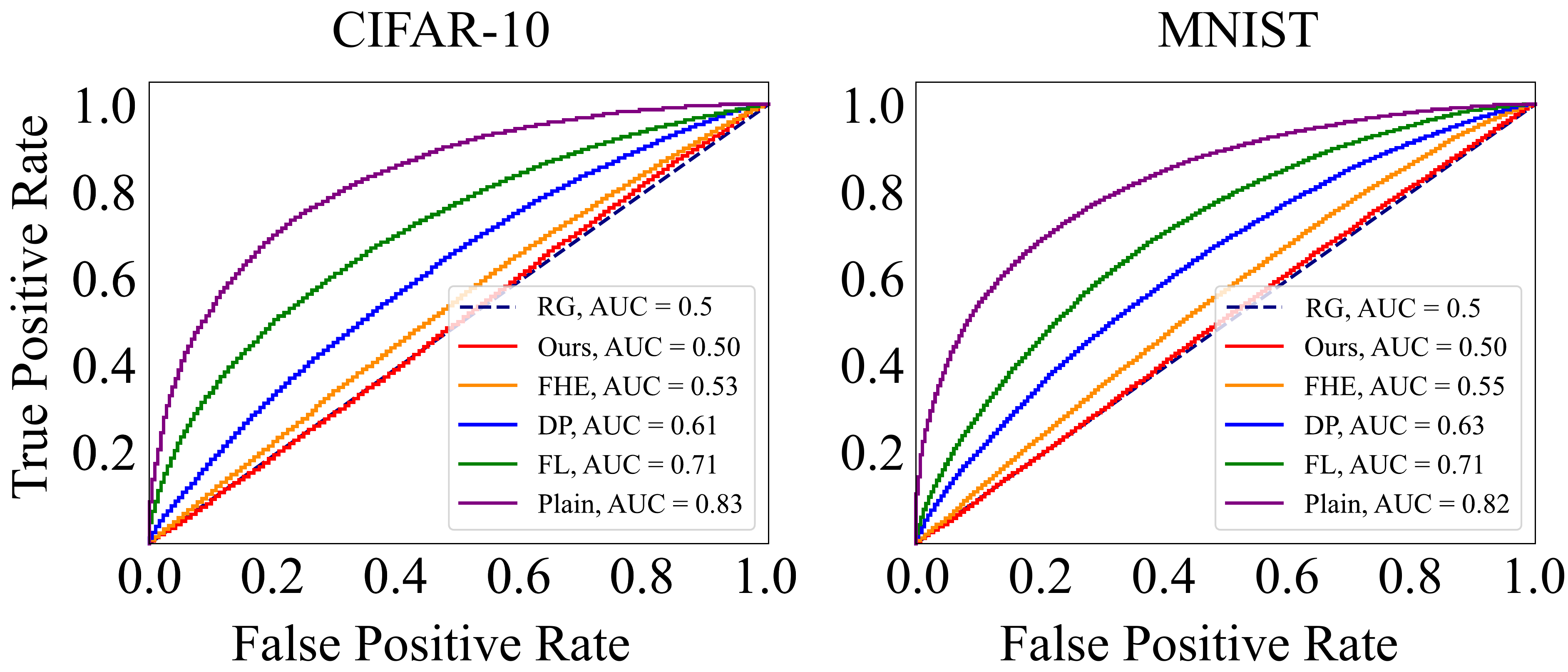}
\caption{ROC results of membership inference attack in \cite{membership-inference-attacks} on the CIFAR-10 and MNIST datasets. The red curve represents ROC trained by \shuffle{}. RG means random guess.}
\label{fig: membership inference attacks results}
\end{figure}

As shown in \autoref{fig: gan attacks results}, GAN methods can reconstruct many facial features from the original data model. It is similar for models with added DP noise in their weights. In the original federated learning approach, client-side trained intermediate updates are uploaded to the server, and facial features from client data can also be extracted. Our method not only protect dataset, but also effectively defends against such attacks, making it nearly impossible to recover identifiable features.

\begin{table*}[t]
\centering
\caption{Accuracy of trained models with different datasets.}\label{tab: accuracy loss}
\begin{threeparttable}
\begin{tabular}{c|ccccc|ccccc|cccc}
\toprule
Model      & \multicolumn{5}{c}{MNIST}      & \multicolumn{5}{c}{CIFAR-10}          & \multicolumn{4}{c}{ImageNet-100\tnote{5} }       \\
\midrule
Method     & Plain & \projname{} & DP & FHE\tnote{4} & \added{InstaHide~\cite{instahide}}       & Plain & \projname{} & DP & FHE\tnote{4} & \added{InstaHide}               & Plain & \projname{} & DP & \added{InstaHide}\\
\midrule
Privacy    & \XSolidBrush & \checkmark & $\bigcirc$\tnote{1} & \checkmark & \minor{$\bigcirc$\tnote{7}} & \XSolidBrush & \checkmark & $\bigcirc$\tnote{1} & \checkmark & \minor{$\bigcirc$\tnote{7}} & \XSolidBrush & \checkmark & $\bigcirc$\tnote{1} & \minor{$\bigcirc$\tnote{7}}\\
ViT-B~\cite{vit} & 99.87\% & 99.14\% & -\tnote{2} & -\tnote{4} & \minor{9.97\%} & 98.63\% & 92.35\% & -\tnote{2} & -\tnote{4} & \minor{10.03\%} & 74.54\% & 72.98\% & -\tnote{2} & \minor{1.03\%}\\
Swin-T~\cite{swint} & 98.72\% & 98.70\% & -\tnote{2} & -\tnote{4} & \minor{10.16\%} & 92.33\% & 85.73\% & -\tnote{2} &  -\tnote{4} & \minor{9.82\%} & 84.80\% & 81.12\% & -\tnote{2} & \minor{0.10\%}\\
ResNet~\cite{resnet} & 99.27\% & 98.81\% & 61.36\%\tnote{3} & -\tnote{4} & \minor{98.79\%} & 97.23\% & 90.15\% & 62.74\%~\tnote{3} & 87.84\%\tnote{6} & \minor{90.04\%} & 90.34\% & 83.78\% & 60.82\%\tnote{3} & \minor{31.08\%}\\
ShuffleNet~\cite{shufflenet} & 98.93\% & 97.19\% & 58.91\%\tnote{3} & -\tnote{4} & \minor{96.27\%} & 86.87\% & 84.07\% & 52.06\%\tnote{3} & -\tnote{4} & \minor{84.97\%} & 85.34\% & 83.64\% & 48.75\%\tnote{3} & \minor{29.78\%}\\
MobileNet~\cite{mobilenet} & 97.21\% & 97.20\% & 51.48\%\tnote{3} & -\tnote{4} & \minor{97.13\%} & 81.37\% & 81.02\% & 59.77\%\tnote{3} & -\tnote{4} & \minor{75.53\%} & 82.94\% & 81.38\% & 48.57\%\tnote{3} & \minor{30.94\%}\\
VGG~\cite{vgg} & 99.51\% & 98.12\% & 69.34\%\tnote{3} & \tnote{4} & \minor{98.05\%} & 82.64\% & 82.63\% & 53.89\%\tnote{3} & 84.76\%\tnote{6} & \minor{82.57\%} & 74.02\% & 73.88\% & 43.56\%\tnote{3} & \minor{1.38\%}\\
\bottomrule
\end{tabular}
\begin{tablenotes}
    \item[1] Here the $\sigma$ of Gaussian noise is $50$. As described in \autoref{sec: Limitation of DP on Releasing Visual Datasets}, DP usually protect data membership, while it has limitation to protect releasing visual dataset.
    \item[2] DP does not support the multi-head attention mechanism in the Transformer architecture.
    \item[3] As DP does not support the batch norm layer in CNNs, it may influence accuracy.
    \item[4] Due to too long execution time, we only provide data of accuracy that we can find in public papers.
    \item[5] As ILSVRC's rule, we use Top-5 accuracy of models on the ImageNet dataset.
    \item[6] Due to too long execution time, results of FHE are cited by \cite{9460962} with VGG-7 and ResNet-20.
    \minor{\item[7] The privacy of InstaHide~\cite{instahide} has been previously called into question~\cite{attack-on-instahide}.}
\end{tablenotes}
\end{threeparttable}
\end{table*}

\begin{table*}[t]
    \centering
    \caption{Throughput (images per second) of different methods on different datasets.}\label{tab: throughput}
    \begin{threeparttable}
    \begin{tabular}{m{0.08\textwidth}<{\centering}m{0.05\textwidth}<{\centering}m{0.11\textwidth}<{\centering}m{0.11\textwidth}<{\centering}m{0.11\textwidth}<{\centering}m{0.11\textwidth}<{\centering}m{0.11\textwidth}<{\centering}m{0.1\textwidth}<{\centering}}
        \toprule
        Method      & Privacy    & ShuffleNet \cite{shufflenet}             & VGG \cite{vgg}             & ResNet \cite{resnet}          & MobileNet \cite{mobilenet}	        & ViT-B \cite{vit}   & Swin-T \cite{swint} \\
        \midrule
        Plain       & \XSolidBrush & 1088.9               & 404.7             & 600.2           & 1070.8            & 322.2 & 472.3 \\
        DP \cite{differential-privacy}          & $\bigcirc$\tnote{1} & 212.3 [-80.5\%]    & 66.1 [-83.7\%]   & 187.8 [-68.7\%]   & 92.1 [-91.4\%]   & -\tnote{2} & -\tnote{2}\\
        FL\tnote{3} \cite{mcmahan2017communication} & $\bigcirc$\tnote{1} & 291.8 [-73.2\%]    & 385.2 [4.8\%]    & 565.1 [5.8\%]  & 1008.7 [-5.8\%] & -\tnote{4} & -\tnote{4}\\
        DP + FL\tnote{3} \cite{differential-privacy, mcmahan2017communication} & $\bigcirc$\tnote{1} & 10.9 [-99.0\%]    & 2.0 [-99.5\%]    & 7.4 [-98.8\%]  & 12.8 [-98.8\%] & -\tnote{4} & -\tnote{4}\\
        FHE\tnote{5}  \cite{9460962}         & \Checkmark   & 0.006[-99.9\%]           & 0.0009 [-99.9\%]                 & 0.0005[-99.9\%]                 & 0.005 [-99.9\%]             & 0.000074 [-99.9\%] & 0.00045 [-99.9\%]\\
        \minor{InstaHide~\cite{instahide}} & \minor{$\bigcirc$\tnote{6}} & \minor{1087.1 [-0.17\%]}       & \minor{399.2 [-1.4\%]}     & \minor{594.1 [-1.0\%]}     & \minor{1062.3 [-7.9\%]}	 & \minor{315.8 [-2.0\%]} & \minor{458.3 [-3.0\%]}\\
        \projname{}  & \Checkmark   & 1080.3 [-0.8\%]       & 401.9 [-0.6\%]     & 595.4 [-0.8\%]     & 1062.1 [-0.8\%]	 & 319.9 [-0.7\%] & 466.1 [-1.3\%]\\
        \bottomrule
    \end{tabular}
    \begin{tablenotes}
     \item[1] Both DP and AvgFL can only protect partial privacy while training\cite{invisible_fl}.
     \item[2] FL represents AvgFL\cite{mcmahan2017communication}. This method also does not support transformer-based models.
     \item[3] Federated learning employs virtual terminals, hence the network transmission delay can be considered as virtually non-existent.
     \item[4] We employ a Python DP library\cite{diffprivlib} developed by IBM Research. This method does not support transformer-based models.
     \item[5] Due to the excessive runtime of FHE, we calculate the average throughput within limited iterations.
     \minor{\item[6] The privacy of InstaHide\cite{instahide} has been previously called into question~\cite{attack-on-instahide}.}
    \end{tablenotes}
    \end{threeparttable}
\end{table*}

As shown in the figure, GAN methods can accurately reconstruct facial features from the original data model. The same holds true for models with added DP noise in their weights. In the original federated learning approach, client-side trained intermediate updates are uploaded to the server, and facial features from client data can also be extracted. Our method effectively defends against such attacks, making it nearly impossible to recover identifiable features.

\textbf{Membership inference attacks.} We also validate another attack method that specifically targets the training weights. This method, \ie, membership inference, aims to determine whether a given dataset has been involved in the training process of the weights.

\autoref{fig: membership inference attacks results} presents the impact of \cite{membership-inference-attacks} method on MNIST and CIFAR-10 datasets. The purple curve indicates that \cite{membership-inference-attacks} method can accurately infer on model trained by raw dataset, with a high probability, whether a given image belongs to the training set. On the other hand, the orange and red curves closely resemble the RG(Random Guess) curve. Other methods usually give different leakages of membership privacy.

\subsection{Validating Training Performance}

\begin{table}[t]
\centering\caption{\small{Federated learning accuracy of ResNet50 on ImageNet via plain scheme and \projname{} scheme.}}
\label{tab: fl}
\begin{tabular}{cccc}
\toprule
\scriptsize Model                             & \scriptsize Method         &\scriptsize MNIST   & \scriptsize CIFAR10\\
\midrule
\multirow{2}{*}{\scriptsize ResNet \cite{resnet}}           & \scriptsize FL             & \scriptsize 99.28\% & \scriptsize 70.83\%\\
                                  & \scriptsize FL+\projname   & \scriptsize 94.39\% & \scriptsize 66.11\%\\
\midrule

\multirow{2}{*}{\scriptsize VGG \cite{vgg}}              & \scriptsize FL             & \scriptsize 99.48\% & \scriptsize 77.29\%\\
                                  & \scriptsize FL+\projname   & \scriptsize 97.25\% & \scriptsize 76.87\%\\
\midrule

\multirow{2}{*}{\scriptsize MobileNet \cite{mobilenet}}        & \scriptsize FL             & \scriptsize 99.23\% & \scriptsize 75.26\%\\
                                  & \scriptsize FL+\projname   & \scriptsize 97.44\% & \scriptsize 73.11\%\\
\midrule
\multirow{2}{*}{\scriptsize ShuffleNet \cite{shufflenet}}       & \scriptsize FL             & \scriptsize 99.20\% & \scriptsize 72.63\%\\
                                  & \scriptsize FL+\projname   & \scriptsize 97.33\% & \scriptsize 72.08\%\\
\midrule

\multirow{2}{*}{\scriptsize Swin-T \cite{swint}}           & \scriptsize FL             & \scriptsize 95.47\% & \scriptsize 67.53\%\\
                                  & \scriptsize FL+\projname & \scriptsize 93.71\% & \scriptsize 61.37\%\\
\midrule

\multirow{2}{*}{\scriptsize ViT-B \cite{vit}}            & \scriptsize FL             & \scriptsize 92.29\% & \scriptsize 60.03\%\\
                                  & \scriptsize FL+\projname & \scriptsize 87.23\% & \scriptsize 57.70\%\\
\bottomrule
\end{tabular}%
\end{table}

In this section, we evaluate the performance of our proposed approach in terms of accuracy loss and throughput introduced by privacy protection. \minor{We compare our method with baseline models, including data-level representative methods such as methods based on fully homomorphic encryption\cite{9460962}, as well as function-level representative methods like Differential Privacy and Federated Learning.}

\textbf{Accuracy}. To comprehensively assess the impact of our method on the accuracy of the inference model, we utilize open-source datasets with varying input resolutions and color channels, including ImageNet-100, CIFAR10, and MNIST. Furthermore, we consider DNN models of different sizes and complexities, such as ViT-B, Swin-T, ResNet-50, ShuffleNet, MobileNet and VGG16.

The experimental results are presented in \autoref{tab: accuracy loss} are obtained through training and validating the models using the MNIST, CIFAR-10, and ImageNet-100 datasets. 
In general, on small datasets like MNIST, the accuracy loss is typically negligible. On larger datasets like ImageNet-100, our method exhibits significant advantages in terms of accuracy compared to DP. Even when compared to models trained on the original dataset, the accuracy loss of our method remains available. \minor{InstaHide~\cite{instahide} shows an accuracy similar to ours on the CNN model. However, it cannot support the Transfomer-based models well, which can be demonstrated by that both Vit-B and Swin-T show low accuracy in InstaHide. The homomorphic encryption, while preserving the model structure, necessitates the substitution of certain activation functions, leading to a decrease in accuracy. Its performance in terms of accuracy is inferior to our approach, but it is markedly superior to methods such as DP that alter the feature map and data.}

\textbf{Throughput}. \minor{We measured the training throughput of our method on different datasets and models in \autoref{tab: throughput}. Our method involves offline processing of the dataset and does not modify the forward and backward computation processes of the model. Thus, it brings little additional computational overhead, similar to InstaHide~\cite{instahide}. As shown in \autoref{tab: throughput}, our method and InstaHide~\cite{instahide} both achieve comparable throughput to the plain training process. Compared to DP-based method and FHE-based method, our method demonstrates significant performance improvement.} However, it should be noted that in the case of ViT-B and Swin-T, an additional 50\% of epochs were required for convergence, while maintaining the same number of epochs in other CNN models. \minor{The homomorphic encryption method requires the decomposition of underlying operators into supported operations to facilitate cryptographically-based procedures. This implies that data can be safeguarded through cryptographic measures without divulging almost any information. However, this also results in a substantial performance degradation, making it challenging for neural network accelerators, such as GPUs and NPUs, to enhance its computational speed. In terms of throughput, its performance is significantly inferior compared to other benchmarked approaches, even with its encryption-grade privacy.}

\subsection{Validating DNN-Related Paradigms}

In addition to its effectiveness in normal DNN training and inference, \shuffle{} provides possibilities for various vision-based DNN-related tasks, including federated learning and knowledge distillation.

\shuffle{} can be applied to the federated learning framework, improving data privacy. By shuffling the original data on each client before training, \shuffle{} mitigates the risk of sensitive information recovery, even in the event of model parameter leakage. The only concern we need to focus on is whether the accuracy loss caused by \projname{} is acceptable. Accuracy experimental results presented in \autoref{tab: fl} showcase low accuracy loss. These results demonstrate the practical value and effectiveness of \shuffle{} in the context of federated learning.

\renewcommand{\arraystretch}{0.8} 
\begin{table}[t]
    \centering\caption{\minor{Knowledge distillation \cite{hinton2015distilling} accuracy of ResNet50 on ImageNet-100 via different training schemes.}}\label{tab: kd}
    \begin{threeparttable}
     \begin{tabular}{m{0.12\textwidth}<{\centering}m{0.06\textwidth}<{\centering}m{0.06\textwidth}<{\centering}m{0.06\textwidth}<{\centering}m{0.06\textwidth}}
        \toprule
        Model & Top-1 & Top-3 & Top-5 & Top-10 \\
        \midrule
        \textbf{TO}-Resnet50 & 74.55\% & 88.42\% & 92.02\% & 95.23\% \\
        \textbf{TV}\tnote{1}-Resnet50 & 66.45\% & 82.67\% & 87.24\% & 91.78\% \\
        \midrule
        \textbf{SO}\tnote{2}-MobileNetv3 & 21.1\% & 44.0\% & 53.6\% & 62.4\% \\
        \textbf{SV}-MobileNetv3 & 18.9\% & 43.1\% & 53.4\% & 62.4\% \\
        \bottomrule
    \end{tabular}
    \tnote{1}\textbf{TV} is \textbf{T}eacher model with \textbf{V}IM.  ~~~~\tnote{2}\textbf{SO} is \textbf{S}tudent model with \textbf{O}riginal.
    \end{threeparttable}
    \vspace{-1pt}
\end{table}

Furthermore, \shuffle{} is leveraged within the knowledge distillation to ensure the privacy of data during the training process of teacher models, and protect privacy of teacher models during distillation. 

Teacher models are often large and valuable models, which can potentially leak information about the dataset or features used to train them. Our method, as shown in \autoref{fig: gan attacks results} and \autoref{fig: membership inference attacks results}, can protect the privacy of the model. In this section, we measure the potential accuracy loss that \projname{} may introduce during the knowledge distillation process.
The experimental results, as illustrated in \autoref{tab: kd}, demonstrate ResNet50 achieving an accuracy of 97.3\% accuracy, while the distilled MobileNet achieved a commendable accuracy of 62.4\%, which is in close proximity to the accuracy achieved without \shuffle{} processing. This highlights the effectiveness of \shuffle{} within the knowledge distillation framework.

In conclusion, \shuffle{} presents a promising solution to enhance data privacy in various DNN applications, including federated learning and knowledge distillation, while maintaining optimal model performance. It holds great potential in supporting diverse DNN scenarios.

\subsection{Evaluation on Object Detection Tasks}

Considering the underlying principles shared between object detection and image classification, the effectiveness of our proposed method in image classification tasks suggests its potential for performing object detection as well. The experimental results for object detection using our method are presented in \autoref{tab: od}.

According to the results, our encryption method demonstrates unique characteristics in the object detection task. While precision remains close to baseline levels, recall and mean average precision (mAP) show a decline. This performance disparity provides valuable insights into the interaction between the model and the obfuscated data. The relatively high precision suggests that the majority of bounding boxes identified by the model contain true positives. In other words, once the bounding box is established, the model can accurately classify the object within it. This aligns with the strong performance observed in classification tasks, indicating that the encryption method has minimal impact on the model's ability to classify objects within a bounded area.

The decrease in recall and mAP indicates that the model struggles to identify and establish bounding boxes around all instances of objects within the image. This results in a lower recall rate, as many true positives are not detected, and the model's effectiveness across various recall thresholds is diminished, leading to a lower mAP. This behavior may be attributed to the model's reliance on contextual information or background cues in classification tasks, which might be less effective in the object detection context, where precise delineation of object boundaries is required. However, this observation does not contradict the effectiveness of the encryption method in classification tasks. It merely  highlights that while the model excels at accurately classifying objects within established bounding boxes (high precision), further fine-tuning or additional techniques may be necessary to enhance its ability to identify and establish bounding boxes around all relevant objects, thereby improving recall and mAP.

In conclusion, this experiment validates the applicability of our encryption method not only in classification tasks but also have potential in object detection tasks. Although there is a decrease in recall and mAP, the high precision emphasizes its practical value. These results further reinforce our earlier findings and establish the versatility of the proposed encryption method in various computer vision tasks.

\renewcommand{\arraystretch}{0.8} 
\begin{table}[]
\caption{Experimental results on VOC dataset.}\label{tab: od}
\centering
     \begin{tabular}{m{0.12\textwidth}<{\centering}m{0.06\textwidth}<{\centering}m{0.06\textwidth}<{\centering}m{0.06\textwidth}<{\centering}m{0.06\textwidth}}
\toprule
                        \textbf{Model}      & \textbf{Method}          & \textbf{Precision} & \textbf{Recall} & \textbf{mAP@50} \\
\midrule
\multirow{2}{*}{YOLO v5 \cite{yolov5}}      & Plain & 0.601     & 0.534  & 0.562  \\
                              & \projname{}     & 0.602     & 0.415  & 0.441  \\
\midrule
\multirow{2}{*}{SSD \cite{ssd}}          & Plain & 0.631     & 0.594  & 0.504  \\
                              & \projname{}     & 0.556     & 0.418  & 0.372  \\
\midrule
\multirow{2}{*}{EfficientDet \cite{efficientdet}} & Plain & 0.817     & 0.660  & 0.765  \\
                              & \projname{}     & 0.735     & 0.419  & 0.505  \\
\bottomrule
\end{tabular}
\end{table}
\section{Related Work}
As deep learning triggers the knowledge extraction capability from heterogeneous data, privacy preservation has been a significant concern and hot research topic for many years. Although existing works have demonstrated their effectiveness in different scenarios, how to protect the privacy of released visual data while preserve data utility is still an open challenge. We briefly review the related work in two categories.

\subsection{Obfuscation-based Mechanisms}
The dominating privacy-preserving mechanism for visual data is obfuscation, which is to obscure the private information (e.g., blurring \cite{aditya2016pic} or blocking \cite{wang2017scalable}) or add some noise (e.g., \cite{qi2022privacypreserving} and DP-based solutions \cite{278384, dwork2006differential, abadi2016deep, zhu2020private, Luo_2021_CVPR, de2022unlocking}) in the image before releasing it for analysis.

A significant problem in obscuring mechanisms is that they
require perfectly accurate and comprehensive knowledge of
the spatial locations of private information in the image. It is usually costly or even infeasible to semantically define what is private in different scenarios \cite{278384}. Besides, obscuring the private information of the image often leads to very low data utility, making it useless for most vision tasks (like classification, detection, etc.).

Differential Privacy \cite{dwork2006differential, abadi2016deep} is another kind of obfuscation method. It preserves data privacy by adding noise, conforming to a specific distribution (like Laplace or Gaussian), to the original data. The key parameter in differential privacy is the privacy budget, which determines the amount of noise added. The advantage of differential privacy lies in its strict mathematical guarantee of data privacy under various data analysis attacks. This method has been successfully applied in many areas, e.g., social networks and medical information systems. Recently, Zhu, Yu, et al. \cite{zhu2020private} proposed a method avoiding splitting the training dataset and achieves comparable or better accuracy while reducing the privacy loss. Luo, Wu, et al \cite{Luo_2021_CVPR} presented an approach minimizing trainable parameters, achieving commendable performance in extensive experiments on diverse visual recognition tasks. However, despite the competitive performance of differential privacy in many fields, it faces some challenges in computer vision, particularly in releasing image data \cite{machanavajjhala2017differential}. In protecting the privacy of image data, differential privacy needs to add a considerable amount of noise to ensure every pixel in the image is sufficiently protected. However, for  human visual system, even with the addition of substantial noise in the image, we can still recognize the main content of the image. This is because human visual system is highly sensitive to edge and texture information in the image, constituting the principal elements of visual features. Thus, although differential privacy can decrease privacy risk mathematically like data membership, it may fail to effectively prevent humans from extracting useful information from image data, especially when specific areas in the image contain sensitive information. It should be noted that the latest work by Google DeepMind \cite{de2022unlocking} proposes a DP based privacy-preserving methods that can make some NF-model accuracy comparable to our solution. However, it still cannot avoid the aforementioned privacy issues, especially when it comes to the release of the dataset.


The $k$-anonymity algorithm \cite{sweeney2002achieving, gedik2007protecting} can also be regarded as a data obfuscation method that protects data privacy by making records in a dataset consistent on certain attributes. Thus, any record is at least identical to $k-1$ other records in these attributes. Nevertheless, the application of the k-anonymity algorithm also has its limitations, because selecting suitable attributes for consistent processing is not a simple task for complex visual data. 

Federated Learning (FL) \cite{konevcny2016federated, ghosh2020efficient} has also emerged to be a feasible privacy preservation mechanism. It protects the data privacy by locally training a model via private data, and shares the model parameters instead of the original data to share knowledge. Although many studies have pointed out the privacy risks in FL\cite{invisible_fl}, its design philosophy is orthogonal from our solution. In our experiments, we also show that applying our solution to FL, we can provide a stronger protection strength to resist the privacy attacks on uploaded local model parameters.

\subsection{Encryption Based Mechanisms}
There are numerous studies focusing on data privacy protection via encrypting the original data or using Trusted Execution Environment (TEE) to isolate private data computing.

Fully Homomorphic Encryption (FHE) \cite{gentry2009fully, van2010fully} is a cryptographic technique that allows arbitrary calculations on ciphertext without needing to decrypt it first. Through FHE, data owners can encrypt their data and send it to cloud service providers for processing, without worrying about data privacy leakage. However, while FHE is wonderful in theory, we are still facing some challenges in practice. The primary issue is computational complexity. FHE algorithms require a substantial amount of computational resources when executing encryption and decryption operations. In particular, arithmetic operations under FHE often have an exponential level of complexity. It usually bring 10x-10,000x additional time overhead on same hardware. Therefore, FHE might become a bottleneck for large-scale image data processing. When handling large-scale data, the computational complexity of FHE could make computation time and resource consumption excessively large, which limits the application of FHE in the field of image data processing.

Another method involves using TEE \cite{mo2021ppfl, chen2020training}, like Intel's SGX and ARM's TrustZone, which can protect the data being processed at the hardware level, preventing unauthorized access and modification. However, the hardware resources of TEE is usually limited and its compatibility to neural network accelerators (e.g., GPU and NPU) is still far from satisfaction, which significantly impacts the performance of DNN computing.

\minor{\section{Discussion}

We acknowledge that, compared to traditional encryption methods, VisualMixer lacks a formal security proof to prove its effectiveness against heuristic attacks. VisualMixer focuses on reducing the visual recognizability of images while enhancing their usability within DNNs. However, as demonstrated by our experiments, VisualMixer is capable of resisting state-of-the-art image restoration attacks. Unlike Differential Privacy (DP) or InstaHide~\cite{instahide}, VisualMixer does not introduce external noise in image obfuscation. The advantage of such an intrinsic transformation is that it is not susceptible to the attacks based on statistical features.

Moreover, VisualMixer can be compatible with other privacy-preserving methods. In \autoref{sec: evaluation}, we validate the compatibility between VisualMixer and Federated Learning (FL). The experiment results therein illustrate that VisualMixer can mitigate privacy issues caused by gradient leakage in FL with minimal accuracy loss. Also, VisualMixer is orthogonal to Differential Privacy (DP). Thus, VisualMixer can enhance DP by further obfuscating images and feature maps without significant accuracy decrease. Note that, VisualMixer is compatible with FHE and obfuscates data before data encryption. However, as FHE already offers very stringent security guarantees, integrating VisualMixer may not be necessary.

}
\section{Conclusion}
\label{sec: conclusion}

In this paper, we introduced \shuffle{},  an innovative and effective strategy for protecting data privacy in the training and inference processes of DNNs. The primary objective of \shuffle{} is to enhance data privacy while preserving the performance of the model. 
To address the inherent trade-off between privacy and model accuracy in data obfuscation strategies, we introduce a new metric, named VFE, to measure the visual privacy, and propose a non-uniform shuffling strategy \shuffle{} to pre-process the image dataset. \shuffle{} allows for effective data obfuscation while ensuring the output deviation of the model does not exceed a preset threshold. We also devised an optimizer, named ST-Adam, to tackle potential dense gradient issues during training. Extensive experiments demonstrate \shuffle{}'s capability to enhance data privacy without compromising the overall performance of the model. In addition, we have identified the potential of \shuffle{} in both federated learning and knowledge distillation frameworks, showcasing its adaptability. 
In summary, \shuffle{} presents a promising data privacy protection method for secure training and inference within DNNs. Our ongoing and in-depth research and exploration of \shuffle{} aim to unlock further possibilities in data privacy protection and provide more safeguards for secure training and inference.

\section*{Acknowledgment}

This research was supported in part by the National Key R\&D Program of China under Grant No. 2022YFF0604502, the National Natural Science Foundation of China under Grant No. 62122095, 62341201, 62132011, 62072472 and U19A2067, the China Postdoctoral Science Foundation under Grant No. 2022M721827, a grant from Kuaishou Technology, and by a grant from the Guoqiang Institute, Tsinghua University.

\clearpage
\bibliographystyle{IEEEtranSN}
\bibliography{refers}
 
\clearpage
\appendices

\section{VIM-ed Sample}

There are some random samples of plain and VIM-ed images of ImageNet-100 dataset.

\begin{figure}[H]
\centering\includegraphics[width=\linewidth] {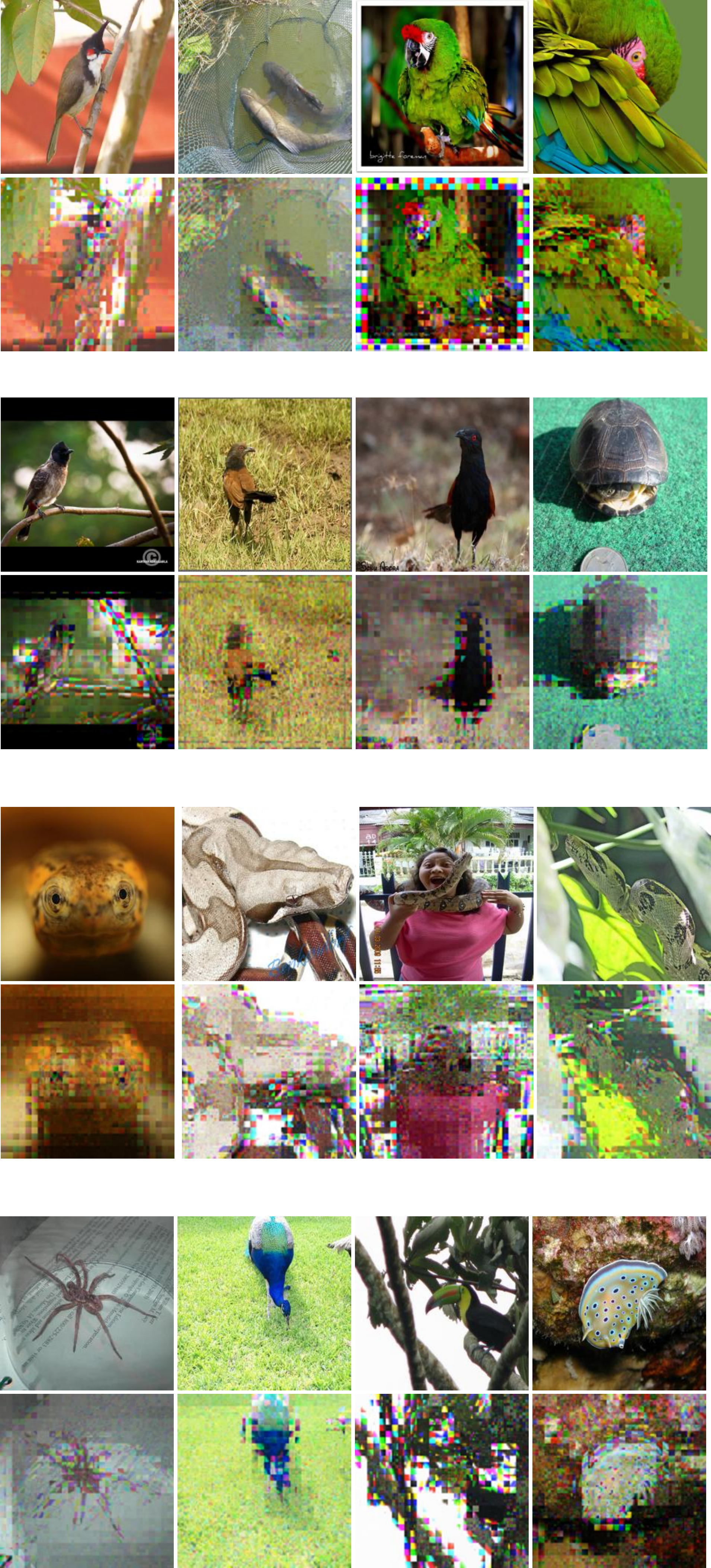}
\end{figure}

\end{document}